\documentclass[prc,twocolumn,superscriptaddress,showpacs,twoside,floatfix]
{revtex4-1}

\usepackage{amssymb,epsfig}
\usepackage{color}
\usepackage{graphicx}
\usepackage{amsmath}
\usepackage{amssymb}
\usepackage{paralist}
\usepackage[percent]{overpic}
\usepackage{rotating}

\newcommand{\thetaBe}{$\theta_{\rm Be}$}	
\newcommand{\Et}{$ E_T $}			
\newcommand{\gPhase}{$\varphi_{02}$}	
\newcommand{\thetaK}{$\theta_{k}$}		
\newcommand{\thetaAlpha}{$\theta_\alpha$}	

\newcommand{\mcm}{$\rm \mu m$}				

\newcommand{\ssd}{SSSD}
\newcommand{\csi}{CsI(Tl)}

\begin{document}


\title{Three-body correlations in direct reactions: Example of $^{6}$Be
populated in $(p,n)$ reaction}

\author{V.~Chudoba}
\email{chudoba@jinr.ru}
\affiliation{Flerov Laboratory of Nuclear Reactions, JINR,  141980 Dubna, Russia}
\affiliation{Institute of Physics, Silesian University in Opava, 74601 Opava, Czech Republic}

\author{L.V.~Grigorenko}
\affiliation{Flerov Laboratory of Nuclear Reactions, JINR,  141980 Dubna, Russia}
\affiliation{National Research Nuclear University ``MEPhI'',
115409 Moscow, Russia}
\affiliation{National Research Centre ``Kurchatov Institute'', Kurchatov
sq.\ 1, 123182 Moscow, Russia}

\author{A.S.~Fomichev}
\affiliation{Flerov Laboratory of Nuclear Reactions, JINR,  141980 Dubna, Russia}

\author{A.A.~Bezbakh}
\affiliation{Flerov Laboratory of Nuclear Reactions, JINR,  141980 Dubna, Russia}

\author{I.A.~Egorova}
\affiliation{Bogoliubov Laboratory of Theoretical Physics, JINR, 141980 Dubna,
Russia}
\affiliation{Department of Physics, Western Michigan University, Kalamazoo, MI 49008, USA}

\author{S.N.~Ershov}
\affiliation{Bogoliubov Laboratory of Theoretical Physics, JINR, 141980 Dubna,
Russia}

\author{M.S.~Golovkov}
\affiliation{Flerov Laboratory of Nuclear Reactions, JINR,  141980 Dubna, Russia}

\author{A.V.~Gorshkov}
\affiliation{Flerov Laboratory of Nuclear Reactions, JINR,  141980 Dubna, Russia}

\author{V.A.~Gorshkov}
\affiliation{Flerov Laboratory of Nuclear Reactions, JINR,  141980 Dubna, Russia}

\author{G.~Kaminski}
\affiliation{Flerov Laboratory of Nuclear Reactions, JINR,  141980 Dubna, Russia}
\affiliation{Heavy Ion Laboratory, University of Warsaw, 02-093 Warszawa, Poland}

\author{S.A.~Krupko}
\affiliation{Flerov Laboratory of Nuclear Reactions, JINR,  141980 Dubna, Russia}

\author{I.~Mukha}
\affiliation{GSI Helmholtzzentrum  f\"{u}r Schwerionenforschung, 64291
Darmstadt, Germany}

\author{E.Yu.~Nikolskii}
\affiliation{National Research Centre ``Kurchatov Institute'', Kurchatov
sq.\ 1, 123182 Moscow, Russia}
\affiliation{Flerov Laboratory of Nuclear Reactions, JINR,  141980 Dubna, Russia}

\author{Yu.L.~Parfenova}
\affiliation{Flerov Laboratory of Nuclear Reactions, JINR,  141980 Dubna, Russia}

\author{S.I.~Sidorchuk}
\affiliation{Flerov Laboratory of Nuclear Reactions, JINR,  141980 Dubna, Russia}

\author{P.G.~Sharov}
\affiliation{Flerov Laboratory of Nuclear Reactions, JINR,  141980 Dubna, Russia}

\author{R.S.~Slepnev}
\affiliation{Flerov Laboratory of Nuclear Reactions, JINR,  141980 Dubna, Russia}

\author{L.~Standylo}
\affiliation{Flerov Laboratory of Nuclear Reactions, JINR,  141980 Dubna, Russia}
\affiliation{Heavy Ion Laboratory, University of Warsaw, Pasteura 5A, Warsaw, Poland}

\author{S.V.~Stepantsov}
\affiliation{Flerov Laboratory of Nuclear Reactions, JINR,  141980 Dubna, Russia}

\author{G.M.~Ter-Akopian}
\affiliation{Flerov Laboratory of Nuclear Reactions, JINR,  141980 Dubna, Russia}

\author{R.~Wolski}
\affiliation{Flerov Laboratory of Nuclear Reactions, JINR,  141980 Dubna, Russia}
\affiliation{Institute of Nuclear Physics PAN, Radzikowskiego 152, 31342 Krak\'{o}w, Poland}

\author{M.V.~Zhukov}
\affiliation{Department of Physics, Chalmers University of Technology, S-41296 G\"oteborg, Sweden}

\date{\today. {\tt d:/latex/6be-s3/6be-s3-5.tex}}

\begin{abstract}
The $^{6}$Be continuum states were populated in the charge-exchange reaction $^1$H($^{6}$Li,$^{6}$Be)$n$ collecting very high statistics data ($\sim 5 \times 10^6$ events) on the three-body $\alpha$+$p$+$p$ correlations. The $^{6}$Be excitation energy region below $\sim 3$ MeV is considered, where the data are dominated by contributions from the $0^+$ and $2^+$ states. It is demonstrated how the high-statistics few-body correlation data can be used to extract detailed information on the reaction mechanism. Such a derivation is based on the fact that highly spin-aligned states are typically populated in the direct reactions.
\end{abstract}

\maketitle


\section{Introduction}
%

The nuclear driplines are defined by instability with respect to particle emission, and therefore the entire spectra of the systems beyond the driplines are continuous. The first emission threshold in the light even systems is often, due to pairing interaction, the threshold for two-neutron or two-proton emission, and therefore one has to deal with three-body continuum.
In certain systems, just beyond the dripline, the continuum of more fragments in the final state can be encountered (e.g.\ $^{7}$H, $^{8}$C, $^{28}$O), so we should speak about few-body continuum. Few-body continuum provides rich information about nuclear structure of ground state and continuum excitations, which is, however, often tightly intertwined with contributions of reaction mechanism. The way to extract this information is to explore the world of various correlations in fragment motions and to look for methods to disentangle contributions of a reaction mechanisms.

Nuclear reactions provide much broader opportunities to study correlations in comparison with nuclear decays. Any reaction has at least one selected direction ---  the direction of a projectile momentum --- and correlations of the reaction products relative to this direction can be studied. This fact is the starting point for a wide-spread method of spin-parity ($J^{\pi}$) identification in the excitation spectrum: experimental angular distribution of the reaction products is compared to Born-type calculations. Such angular distributions can be qualitatively described in terms of transfered momentum $\mathbf{q}$ and transfered angular momentum $\Delta L$ by a simple analytical expression for differential cross-section
\begin{equation}
\frac{d \sigma _{\Delta L}}{d \Omega} \sim  \left| j_{\Delta L} (q r_0) \right|^2 \,,
\label{eq:besselAngDist}
\end{equation}
where $j_{\Delta L}$ is spherical Bessel function and $r_0$ is some typical size of the ``reaction volume''. In spite of quite qualitative character of the dependence Eq.\ \eqref{eq:besselAngDist}, in some cases, it could be sufficient for complete $J^{\pi}$ identification. Applications of such methods are limited by field of \emph{direct reactions}, where the Born-type approximations are robust.

Alternative method of spin-parity identification can be used for a \emph{narrower class of the direct reactions} populating \emph{states in the continuum}. Namely, for direct reactions which can be well described by the pole mechanism [or single diagram with transfer of one species, see Fig.\ \ref{fig:scheme} (a)], where one-step reaction gives dominating contribution. Such a mechanism is widespread at intermediate ($20-70$\,AMeV) and high ($>$70\,AMeV) energies which are commonly used in the modern radioactive ion beam (RIB) research. It selects one exceptional direction in space defined by the vector of the transferred momentum $\mathbf{q}$. In the coordinate frame where $Z$ axis is parallel to $\mathbf{q}$, only a zero projection $\Delta M$ = 0 of the orbital angular momentum $\Delta L$ can be transferred.
\begin{equation}
\{[\Delta \mathbf{L}\times \mathbf{q}]\equiv 0, Z \parallel \mathbf{q}\}
\rightarrow \Delta M=0 \,.
\label{eq:alignment}
\end{equation}
This assumption may be applied only to transfer of spinless particles.
However, in many reaction scenarios the states with $J > 1/2$ are populated with high spin-alignment in the momentum transfer frame even in the case of nonzero spin transfer. For highly aligned states, \emph{decaying via particle emission}, the angular distributions with respect to the axis $Z \parallel \mathbf{q}$ could have very distinctive shape, which can be used for spin-parity identification.

This method was broadly used for spin-parity identification of excited states decaying via emission of (mainly spinless) particles in the past \cite[and Refs.\ therein]{Artemov:1972}. During the last decade, such an approach was applied to exotic neutron- and proton-rich systems beyond the driplines in the experiments at the Flerov Laboratory of Nuclear Reactions at JINR (Dubna, Russia). For example, the interference patterns for broad overlapping states with different $J^{\pi}$ were used for unambiguous spin-parity identification of low-lying $^{9}$He continuous states decaying via $^8$He+n channel \cite{Golovkov:2009}. Analogous method can be used for three-body systems, however in a technically much more complicated manner. The examples of such a $J^{\pi}$ identification in three-body systems can be found for $^{5}$H \cite{Golovkov:2004a,Golovkov:2005} and for $^{10}$He states \cite{Sidorchuk:2012}.

The first results of the experiment studying the $\alpha$+$p$+$p$ correlations in decays of the $^{6}$Be states populated in the $(p,n)$ charge-exchange reaction were published in Ref.~\cite{Fomichev:2012}. The paper was focused on the proof that the observed $^{6}$Be excitation spectrum above $\sim 3$ MeV is dominated by the novel phenomenon -- isovector breed of the soft dipole mode ``built'' on the $^{6}$Li ground state (g.s.). In this work we consider the correlations in the decay of $^{6}$Be states with excitation energy below $\sim 3$ MeV, where the data are dominated by the contributions of the known and well-understood $0^+$ and $2^+$ states of $^{6}$Be. We pursue a sort of an opposite aim to Refs.\ \cite{Golovkov:2009,Golovkov:2004a,Golovkov:2005,Sidorchuk:2012}. We demonstrate that basing on the \emph{known level scheme} it is possible to extract from the three-body correlations the \emph{maximal possible} quantum mechanical information about reaction mechanism (e.g.\ the density-matrix parameters) thus paving the way to its in-depth theoretical studies.

Unit system $\hbar = c = 1$ is used in this work. The article is structured in the following way. First, kinematics notations are given for three-particle correlations detected in a reaction with four particles in final state (Section II). Then  a description of the applied theoretical model is presented in Section III in detail. The experimental setup and conditions are given in Section IV. The data analysis is described in Section V, and the physics discussion and conclusions are in Sections VI and VII, respectively.


\section{Three-body correlations}

\begin{figure}
\begin{center}
\includegraphics[width=0.45\textwidth]{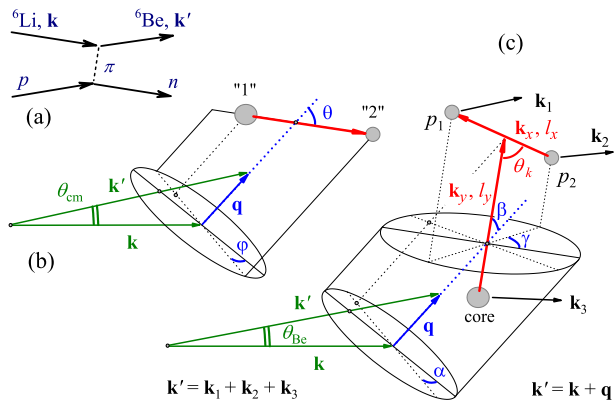}
\end{center}
\caption{Schematic presentation of $^{6}$Be population in charge-exchange reaction and correlations available for particle-unstable states.
(a) Single pole reaction mechanism. Complete kinematics description for correlations following the two-body (b) and three-body (c) decays. In panel (c), the center-of-mass angle $\theta_{\mbox{\scriptsize Be}}$ (green color) describe scattering of $^{6}$Be as a whole, red color shows kinematical variables associated with \emph{internal correlations}, while blue ones are responsible for \emph{external correlations} describing orientation of system as a whole.}
\label{fig:scheme}
\end{figure}

Let us consider three-body correlations obtained in the nuclear reaction
$^6$Li + $p$ $\to$ ($p$ + $p$ + $\alpha$) + $n$. In general, the spin-averaged cross section for a collision $ A + p \to k_1 + k_2 + k_3 + k_n$ of
a projectile $A$ and a proton target $p$, leading to the four fragments
in the final state, can be written in the following way
\begin{eqnarray}
\sigma &=&  {(2\pi)^4 \over v_i} {1 \over \hat{J}_A^2 \hat{J}_p^2}
\sum \int d{\bf k}_1\ d{\bf k}_2\ d{\bf k}_3\ d{\bf k}_n\
 \nonumber \\
 &\times& \delta(E_f - E_i)\ \delta({\bf P}_f - {\bf P}_i)\
 {\mid {\it T}_{fi} \mid}^2,
\label{eq:cross}
\end{eqnarray}
where $E_i = E_p + E_A$, $E_f = E_1 + E_2 + E_3 + E_n + Q$, ${\bf P}_i = {\bf k}_p + {\bf k}_A$, ${\bf P}_f = {\bf k}_1 + {\bf k}_2 + {\bf k}_3 + {\bf k}_{n}$ are the total energies and momenta of all particles before and after collisions, respectively. $Q=-3.70$\,MeV is the separation energy of the nucleus $A$ (the reaction $Q$-value is calculated in respect to the three-body threshold in the final state), $E_j$ is a kinetic energy of particle $j$. The relative incident velocity is $v_i = k_i/ \mu_i$, and $\mu_i = m_p m_A/(m_p+m_A)$ is the reduced mass of the nuclei before collision. Shortcut $\hat{J} = \sqrt{2 J + 1}$ is used in (\ref{eq:cross}), and the summation is over spin projections of all particles before and after collision. In the ($p$ + $A$) center-of-mass (c.m.) coordinate frame ${\bf P}_i = 0$, ${\bf k}_A = -{\bf k}_p = {\bf k}$, $E_i$ = $k^2/2\mu_i$. Our prime interest is in studies of nuclear systems consisted out of the three particles $k_1$, $k_2$ and $k_3$. Then, the fragment relative motion in three-body continuum can be described by two relative Jacobi momenta $\mathbf{k}_x$ and $\mathbf{k}_y$ and the c.m.\ momentum $\mathbf{k}'$ of the three particles
\begin{eqnarray}
{\bf k}_{x} &=& \mu_x \left( {{\bf k}_1 \over m_1} - {{\bf k}_2
\over m_2} \right), \quad \quad \, \mu_x = { m_1 m_2 \over m_{12} }
\nonumber \\
{\bf k}_{y} &=& \mu_y \left( {{\bf k}_1+{\bf k}_2 \over m_{12} } -
{{\bf k}_3 \over m_3} \right), \mu_y = { m_{12} m_3
\over  m_{123} } \nonumber \\
{\bf P}_f &=& 0, \quad \quad {\bf k}' = {\bf k}_1 + {\bf k}_2 + {\bf k}_3 =
-{\bf k}_4
\label{eq:jac}
\end{eqnarray}
where $m_{12}$ = $(m_1 + m_2)$, $m_{123}$ = $(m_1 + m_2+ m_3)$.
For each three-body decay event, the Jacobi momenta $\mathbf{k}_{x}$ and $\mathbf{k}_{y}$ define the decay plane. The \emph{internal correlations} of the fragments [shown by red color in Fig.~\ref{fig:scheme} (c)] are defined within this plane while \emph{external correlations} [blue colored in Fig.~\ref{fig:scheme} (c)] defines orientation of this plane with respect to the reaction plane [green colored in Fig.~\ref{fig:scheme} (c)], which is fixed by the initial $\mathbf{k}$ and final $\mathbf{k}'$ c.m.\ momenta.

Internal three-body correlations for the Jacobi momenta $\mathbf{k}_{x}$ and $\mathbf{k}_{y}$ are conveniently described by two parameters $\{\varepsilon, \theta_k \}$ in the following way:
\begin{eqnarray}
\varepsilon & = & E_x/E_T ,\quad \cos(\theta_k)=(\mathbf{\hat{k}}_{x} \cdot
\mathbf{\hat{k}}_{y})   \,  \nonumber \\
E_T & = & E_x+E_y = k^2_x/2\mu_x + k^2_y/2\mu_y \,.
\label{eq:corel-param}
\end{eqnarray}
The three-body decay energy $E_T$ fixes only a total phase volume accessible for the three fragments, and the fragment kinetic energies have continuous distributions within this volume. In addition, two relative orbital angular momenta $l_x$ and $l_y$, corresponding to the ${\bf k}_x$ and ${\bf k}_y$ momenta, characterize their motion. Since in the $^{6}$Be two fragments are identical protons, only two different distinguishable Jacobi coordinate systems exist. One, labeled ``T'', corresponds to the case when particles 1 and 2 are protons with relative momentum ${\bf k}_x$, while particle 3 is the $\alpha$-particle. In the second case, called ``Y'', the relative momentum ${\bf k}_x$ is defined by the proton with index 1 and $\alpha$-particle with index 2, while the other proton has index 3. The variables $\{\varepsilon, \theta_k \}$ depend on the Jacobi systems while the energy $E_T$ is invariant, i.e., independent from this choice. The representations ``T'' and ``Y'' are equivalent. In spite of this, we use both systems, since certain aspects of correlation can be better revealed in one of them. For example, in the ``T'' system the parameter $\varepsilon$ describes the energy correlation between two protons, while in the ``Y'' it is connected with core-$p$ energy correlations in $^{6}$Be. It is convenient to distinguish parameters $\{\varepsilon_T, \theta_k^{(T)}\}$ and $\{\varepsilon_Y, \theta_k^{(Y)}\}$ in Jacobi ``T''- and ``Y''-systems, respectively.

External correlations describe the orientation of the three-body decay plane relative to the selected direction. We choose this direction along the transferred momentum $\mathbf{q}$ = $\mathbf{k}'$ - $\mathbf{k}$ lying in the reaction plane. In this case the external correlations are three Euler angles labeled as $\{ \alpha, \beta, \gamma \}$ in Fig.~\ref{fig:scheme} (c). This choice has an advantage over other possibilities in the case of one-step reaction mechanism domination. In such a case the particle transfer can be described in a single-pole approximation and there should be rotational invariance with respect to vector $\mathbf{q}$. The decay dynamics will be independent from the angle $\alpha$. This is a manifestation of the so-called Treiman-Yang criterion for the dominance of a single-pole mechanism of direct reactions \cite{Treiman:1962,Shapiro:1965}. This criterion is an important and  useful tool to check the assumption on a reaction mechanism providing its \emph{necessary condition}, and the experimental data can be tested on agreement to it. Concerning other external parameters, the dependence on the orientation angle $\beta$ for three-body decays is easy to interpret, as will be shown below, while an interpretation of the $\gamma$ parameter is not straightforward.

In the end of this section we briefly summarize differences between two- and three-body correlations observed in the decays and in the reactions. On the one hand, the orientation of the system as a whole is assumed to be  isotropic in decays, as far as the system has ``forgotten'' how it was populated. On the other hand, reactions may have several selected directions in the space. The one is the beam direction. For direct reactions with single-pole mechanism there is, as mentioned above, another important direction: the transferred momentum $\mathbf{q}$. So, two following features for the decays and reactions with subsequent two-body and three-body decay can be emphasized:

\noindent (i) Two-body decay is characterized just by two parameters: energy and width of the state.  In contrast, for description of three-body decays, except the energy and width of the state, we need additional parameters, so-called \emph{internal} correlation parameters $\varepsilon$ and $\cos(\theta_k)$.

\noindent (ii) Population of spin-aligned states is common for nuclear reactions. Two additional parameters related to orientation $\{ \theta,\varphi \}$ are needed to describe the two-body decay of aligned state. These are spherical angles for the decay momentum in Fig.~\ref{fig:scheme} (b). In contrast, for description of three-body decays, we need three \emph{external} correlations parameters. Euler angles $\{\alpha,\beta,\gamma \}$ connected with three-body decay plane are convenient to use, see Fig.~\ref{fig:scheme} (c). The angle $\alpha$ is analogous to the angle $\varphi$ in two-body decay, and decay dynamics is independent from it for direct reactions we are interested in. The angle $\beta$ describes an orientation of the decay plane relative $\mathbf{q}$ and is analogous to the angle $\theta$ which describes the direction of decay in two-body case.

\section{Theoretical model}
\label{sec:theory}

The transition matrix element in Eq.\ (\ref{eq:cross}) for the $^1$H($^{6}$Li,$^{6}$Be)$n$ reaction includes all interaction dynamics and is given in prior representation by
\begin{eqnarray}
{\it T}_{fi} &=& \langle \Psi^{(-)}_{M_{p_1}, M_{p_2}, M_n}({\bf k}_x, {\bf k}_y, {\bf k}_f) \mid  \nonumber \\
& & \sum_{i} V_{pi} \mid \Psi_{M_A}, \chi_{M_p}({\bf k}_i) \rangle
\label{eq:tmatr} \\
{\bf k}_f &=& \mu_f \left( {{\bf k}' \over m_{123}} -
{{\bf k}_4 \over m_4} \right),\
\mu_f = {m_4\ m_{123} \over m_4+m_{123} }  \nonumber
\end{eqnarray}
where ${\bf k}_f$ is the relative momentum between c.m. of the $^6$Be nucleus and neutron, $M_i$ denotes the spin projection of the i-th particle, $\Psi_{M_A}$ is the ground state wave function of the $^6$Li nucleus, $\chi_{M_p}({\bf k}_i)$ is a plane wave describing relative motion of proton target and c.m.\ of the $^6$Li nucleus, $\sum_{i} V_{pi}$ is composed of effective nucleon-nucleon interaction $V_{pi}$ between proton target and projectile valence nucleons (marked as $i$). Charge-exchange interaction with $\alpha$-core should not lead to a population of the three-body continuum. The $\Psi^{(-)}_{M_{p_1}, M_{p_2}, M_n}({\bf k}_x, {\bf k}_y, {\bf k}_f)$ is the exact continuum wave function describing relative motion of the four final particles with ingoing-wave boundary conditions. To get  $\Psi^{(-)}$ one has to solve equations of the Faddeev-Yakubovsky type, taking into account the complex nature of the constituents. An exact solution has not been feasible up to now, and therefore approximate methods are required. We make approximations at the level of the reaction mechanism but the three-body structure of the involved nuclei is treated in a consistent way.

At low excitation energies of the $^6$Be nucleus, the relative velocities of the
$^6$Be fragments are small and are restricted kinematically by the $E_T$. It means that interactions between these fragments have to be taken into account. But if collision is relatively fast and a one-step processes are dominated, then a reasonable approximation for $\Psi^{(-)}$ is the following factorization
\begin{eqnarray}
\langle \Psi^{(-)}_{M_{p_1}, M_{p_2}, M_n}({\bf k}_x, {\bf k}_y, {\bf k}_f) \mid \quad & \simeq & \nonumber \\
\langle {\chi}_{M_n}^{(-)}({\bf k}_f),
\Psi^{(-)}_{M_{p_1}, M_{p_2}} ({\bf k}_x,\ {\bf k}_y) \mid \,, & &
\label{eq:psi}
\end{eqnarray}
where $\Psi^{(-)}_{M_{p_1}, M_{p_2}} ({\bf k}_x,\ {\bf k}_y)$ is a continuum three-body wave function of the $^6$Be system with excitation energy $E_T$. $\chi_{M_n}^{(-)}({\bf k}_f)$ is a distorted wave describing relative motion between c.m.\ of the $^6$Be and neutron, which depends on the respective relative coordinate between their center of mass.

If fragments are detected in coincidence, a number of various correlations can be obtained. The exclusive cross section
\[
d^8 \sigma/d\hat{\bf k}_f \, d \hat{\bf k}_x \, d \hat{\bf k}_y \, d \varepsilon \, d E_T\,,
\]
contains the maximum possible information about the nuclear structure and reaction dynamics that can be extracted from a three-body  breakup induced by the collision of two unpolarized nuclei (projectile and target). Exploration of this cross section is quite a challenge both experimentally (huge statistics is demanded) and theoretically, because it involves too many independent variables for transparent analysis. Integrating out some unobserved degrees of freedom brings us to less-exclusive (increasingly inclusive) cross sections. Any integration over a dynamical variable, within its full range of variation, washes out the correlations defined by this degree of freedom. Cross sections after integration become less and less informative, but are simultaneously more suitable for theoretical modeling.
On the other hand, often not all the particles produced by reaction are measured by detectors.
Depending on the geometry of experimental installation and the efficiency of particle registration, some fragments avoid the measurements. Thus, for a proper comparison of theoretical calculations with experimental data, the integration over some unobserved degrees of freedom should be done not within a full range of variation but taking into account response of the experimental setup. Practical way to perform this task is to use the Monte-Carlo simulation of the reaction events. This allows to make an additional simplification in theoretical treatment of the reaction dynamics, namely to substitute the distorted wave $\chi_{M_n}^{(-)}({\bf k}_f)$ by a plane wave. Then, the product of two plane waves in the transition matrix element (\ref{eq:tmatr}) is reduced to the plane wave which depends on the transferred momentum $\mathbf{q}$
\[
\chi_{M_n}^{(-)\star}({\bf k}_f)\, \chi_{M_p}({\bf k}_i) \simeq
\exp[- ({\bf q} \cdot {\bf R})] \, \mid 1 / 2, M_p \rangle
\langle 1 / 2, M_n \mid \,.
\]
Finally, we treat the motion between c.m.\ of colliding systems within the
plane wave approximation (PWA) but the three-body decay dynamics is considered
in a full complexity by taking into account all interactions between fragments.
The disadvantage of such a treatment is that we can not calculate absolute contributions to cross sections from excitations with different values $J^{\pi}$.
However, relative contributions from possible excitation modes leading to the excitation with the fixed value of $J^{\pi}$ can be calculated.
The absolute weights of different $J^{\pi}$ excitations are restored by fitting to the experimental data.
Hereby we remedy our simplified plane wave treatment of the reaction dynamics.

The Hyperspherical Harmonics (HH) method is used for calculations of the three-body continuum wave function. The $\alpha$+$p$+$p$ wave function (WF) of $^{6}$Be with outgoing asymptotics with fixed total momentum $J$ and its projection $M$ is obtained from solution of the Schr\"{o}dinger equation with the source term
\begin{eqnarray}
(\hat{H}_3-E_T)\Psi_3^{JM(+)} = \hat{\mathcal{O}} \, \Psi_{^6\text{Li}}^{J'M'} \,,
\label{eq:shred} \\
\hat{H}_3 = \hat{T}_3 + V_{12}(\mathbf{r}_{12}) + V_{23}(\mathbf{r}_{23}) + V_{31}(\mathbf{r}_{31})\,. \nonumber
\end{eqnarray}
The wave function $\Psi_3^{JM(+)}$ is linked with the wave function $\Psi^{(-)}_{M_{P_1}, M_{P_2}}(\bf{k}_x, \bf{k}_y)$  in Eq. (7) by the reversal of time which involves reversing the linear momenta ($\bf{k}_x$ and $\bf{k}_y$) and direction of the spin rotation ($M_i \rightarrow -M_i$). The effective charge-exchange interaction between projectile and target nucleons with Gaussian formfactor is used
\begin{eqnarray}
\hat{V}(r) & = & V_0 \left[c_{ivs}  + c_{ivv} \left(\mathbf{\sigma}^{(1)} \cdot \mathbf{\sigma}^{(2)} \right) \right] \nonumber \\
& \times & \left(\mathbf{\tau}^{(1)} \cdot \mathbf{\tau}^{(2)} \right) \exp \left[- r^2/r_0^2 \right] \,,
\label{eq:wig-pot}
\end{eqnarray}
where coefficients $c_{ivs}$ and $c_{ivv}$ define the strength of isovector-scalar and isovector-vector couplings. For such an interaction the transition operator in \eqref{eq:shred} is given in the PWA by an analytical expression
\begin{eqnarray}
\hat{\mathcal{O}} \sim \sum \nolimits _i f_l(q,r_i) \left[c_{ivs}  + c_{ivv}
\sigma_{\mu}^{(i)} \right] \tau_{-}^{(i)} Y_{lm}(\hat{\mathbf{r}}_i) \,, \nonumber \\
f_l(q,r_i) = V_0 \, r_0^3 \, \sqrt{2} \, \pi^2 \exp[-(qr_0/2)^2] \, j_l(qr_i),
\label{eq:c-e-oper}
\end{eqnarray}
where index $i$ numbers the two valence nucleons. Such, relatively simple, choice allowed us to reproduce well the angular distributions of $^{6}$Be in Ref.~\cite{Fomichev:2012}.

The exclusive cross section of the direct reaction populating three-body continuum is, in general, an eight-folded differential.
In our specific case, when an orientation of the reaction plane does not play a role, we work with a seven-folded differential cross sections and represent it by using the hyperspherical energy variables as follows
\begin{eqnarray}
\dfrac{d^7 \sigma}{d q \, dE_T \, d \Omega_{\varkappa}} = \sum_{SM_S}
\sum_{JM,J'M'} \rho^{J'M'}_{JM}(q, E_T) \nonumber \\
\times A^{\dagger}_{JMSM_S}(E_T,\Omega_{\varkappa})
\,A_{J'M'SM_S}(E_T,\Omega_{\varkappa})\, ,
\label{eq:rho-matr}
\end{eqnarray}
where $\rho^{J'M'}_{JM}$ is a density matrix and $A_{JMSM_S}$ are three-body amplitudes depending on the $^{6}$Be excitation energy $E_T$ and the five-dimensional hyperspherical ``solid angle''
\[
\Omega_\varkappa = \{ \theta_{\kappa}, \hat{\bf k}_x, \hat{\bf k}_y \},
\]
where $\tan (\theta_{\kappa})$ = $k_x/k_y$. Here, the slow motion of fragments $\alpha$+$p$+$p$ at low excitation energies is described by $A_{JMSM_S}$ amplitudes and the state alignments are contained in the $\rho^{J'M'}_{JM}$.

Note the dependence of the density matrix $\rho^{J'M'}_{JM}$ on the energy $E_T$ and the absolute value of $q$.
In general case there should be a dependence on $\mathbf{q}$, but the azimuthal angle of $\mathbf{q}$ relative to the beam direction is defined by $q$ and $E_T$ for transfer reactions.
Also note that the amplitudes $A_{JMSM_S}$ explicitely depend on $E_T$ and $\Omega_{\varkappa}$, though it can be seen in Eq.\ (\ref{eq:c-e-oper}) that there is also implicit dependence on $q$.
The expression (\ref{eq:rho-matr}) is somewhat different from that stated in Ref.\ \cite{Grigorenko:2012} as far as we explicitly provide summation over spin variables $\{S,M_S\}$ of the three-body channel, which are not measurable (at least in foreseen realistic experimental scenarios).


\section{Experiment}

\begin{figure}[tb]
\centering
\begin{overpic}[width=0.9\linewidth]{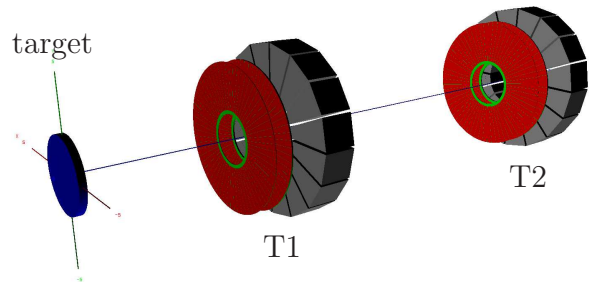}
  \put(43,5){\large T1}
  \put(85,17){\large T2}
  \put(0,40){\large target}
\end{overpic}
\caption{Schematic view of the detector system employed for registration of the $^1$H($^{6}$Li,$^{6}$Be)$n$ reaction products. The origin of the left-handed laboratory frame is in the centre of the hydrogen gas target (blue cylinder), each of two identical telescopes (T1 and T2) consists of two position-sensitive Si detectors (red) and the array of \csi\ crystals (grey).}
\label{fig:setupBe}
\end{figure}

The experiment was performed in the Flerov Laboratory of Nuclear Reaction, Joint Institute for Nuclear Research with the use ACCULINNA setup at U-400M cyclotron \cite{Fomichev:2012}. To carry out high efficiency correlation measurements in the charge-exchange reaction $^1$H($^{6}$Li,$^{6}$Be)$n$, maximal possible statistics of three-particle $\alpha$+$p$+$p$ coincidences were desired. This condition required the detection of at least two particles by one of telescopes (see Fig.~\ref{fig:setupBe}) and the employment of a sophisticated experimental trigger and following data analysis.

The 47 AMeV $^{6}$Li beam was produced by the cyclotron U-400M and injected into ACCULINNA facility \cite{Rodin:1997}. The beam energy was reduced to 35 AMeV using a carbon degrader and delivered to the well shielded experimental room, located behind a 2 m thick concrete wall, where the background produced by cyclotron is considerably suppressed.

Experimental target and detectors setup were placed in stainless steel vacuum reaction chamber pum\-ped out to a stationary pressure of $\sim10^{-6}$\,mbar. The beam was focused on experimental target by means of lead diaphragm positioned between two ionization chambers which compare the beam intensity before and after beam passage through the diaphragm. The beam with intensity of about $3\times 10^7$ $\text{s}^{-1}$ was focused to a $\sim 3$ mm (FWHM) spot in the target plane and an energy spread better than 0.6\% was achieved.

The detector array used for registration of the reaction products and a cryogenic hydrogen target  are shown schematically in Fig.\ \ref{fig:setupBe}. The 4 mm thick target cell was equipped with 6\,\mcm\ stainless steel entrance and exit windows. For the sake of the heat shielding this cell was embedded in a protective volume supplied with 2\,\mcm\ windows of mylar coated with aluminum. The target geometry allowed to detect reaction products emitted in downstream direction with full opening angle of $90^{\circ}$. The target cell was filled with hydrogen gas at a pressure of 3\,bar and cooled down to 35\,K. The difference between the pressure in target cell and the vacuum chamber caused the inflation of steel windows to lenticular form and resulting at maximal target thickness of 6 mm.

Reaction products were measured by two identical annular telescopes T1 and T2, see Fig.\ \ref{fig:setupBe}. Each telescope consisted of two position-sensitive silicon detectors and an array of 16 trapezoid \csi\ crystals coupled with individual S8650 Si-photodiodes. The first double-sided silicon strip detector (DSSD), 300\,\mcm\ thick, had 32 sectors on the front side and 32 rings on the back side. The second layer was made of a single-sided silicon strip detector (SSSD) 1 mm thick, segmented into 16 sectors. The inner and outer diameters of the sensitive area of silicon detectors were 32\,mm and 82\,mm, respectively. The inner diameter of the silicon wafer was 28\,mm. The assembly of \csi\ crystals, 19 mm thick, had the inner and outer diameters of 37\,mm and 97\,mm, respectively. Dead layers of Si detectors were measured using $\alpha$-source, those of \csi\ detectors were estimated by MC simulations.

\begin{figure}[tb]
\centering
\includegraphics[width=\linewidth]{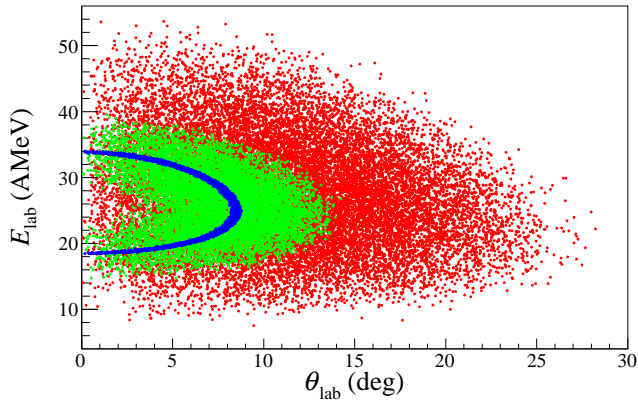}
\caption{Kinematic plot of the $^1$H($^{6}$Li,$^{6}$Be)$n$ reaction and $^{6}$Be decay obtained by MC simulation of the ground and the first excited state population. The beam direction coincides with $Z$-axis in laboratory frame. $\theta_{\rm lab}$ is the polar angle in laboratory frame, $E_{(\rm lab)}$ is the kinetic energy of the reaction and decay products in the laboratory frame. The $^{6}$Be c.m.\ is shown by blue dots, $\alpha$ by green dots and protons by red dots.}
\label{fig:kinematics2be}
\end{figure}

Overall thickness of each telescope was sufficient to stop all products of the investigated process with well-defined identification. DSSDs were intended to measure energy loss $\Delta E$ of particles (with the threshold of $\sim$300\,keV) and the positions of their hits.
SSSD and \csi\ detectors served for measurement of the remaining particle energy deposit. Moreover, signals from \ssd\ were branched to fast time electronic circuit and used for formation of the trigger. The telescopes T1 and T2 were placed 91 mm and 300 mm downstream the target, respectively. Under an assumption that the reaction occurred in the center of the target, the T1 and T2 angular ranges in laboratory frame were  $9.9^{\circ}-24.2^{\circ}$ and $3.1^{\circ}-7.8^{\circ}$, respectively, see Fig.~\ref{fig:kinematics2be}.


\section{Data analysis}
\label{sec:danalysis}

\begin{figure}[tb]
\centering
\includegraphics[width=\linewidth]{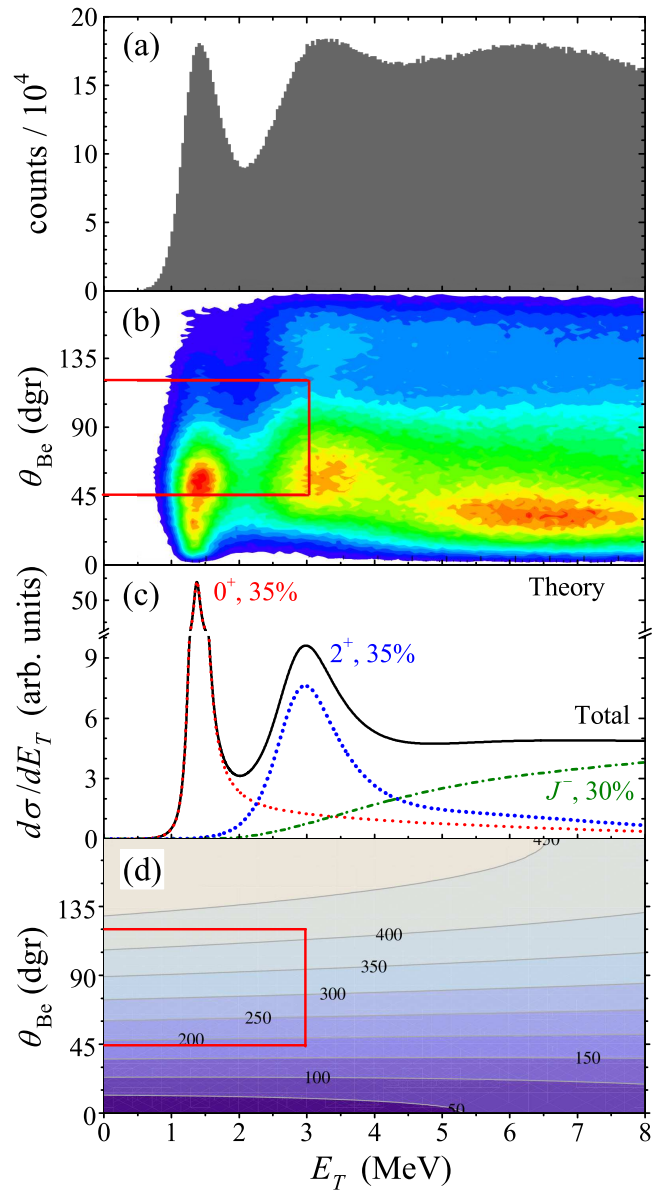}
\caption{Experimental data for $^1$H($^{6}$Li,$^{6}$Be)$n$ reaction. Panel (a) shows the $^{6}$Be integral invariant mass spectrum measured in the whole range of \thetaBe\ angle. Panel (b) shows the same data presented  on the $\{E_T,\theta_{\text{Be}} \}$ plane. Panel (c) shows theoretical spectra with different $J^{\pi}$ and their sum (black solid curve) fitting the data of panels (a) and (b). Panel (d) shows contour plot of the transfered momentum (in MeV/c) on the $\{E_T,\theta_{\text{Be}} \}$ plane. Red rectangle in panels (b) and (d) shows the region of interest for this work.}
\label{fig:IMspectrum}
\end{figure}

As a result of the experiment the $^{6}$Be energy spectra shown in Figs.\ \ref{fig:IMspectrum} (a) and (b) were obtained. The spectrum presented in Fig.\ \ref{fig:IMspectrum} (a) consists of two prominent peaks related to the population of the ground $0^+$ and the first excited $2^+$ states superimposed on the broad continuum. The width of the ground state peak demonstrates overall instrumental resolution. This is a typical picture which has been seen in a number of earlier observations (e.g., see Refs.\ \cite{Tilley:2002,Yang:1995,Guimaraes:2003}) where the low-energy $^6$Be spectrum was populated in charge-exchange reactions. Those results were based on the measurement of the missing mass spectra, and their treatment was often related to the analysis of the excitation spectrum and sometimes its angular behaviour. The detection of three $^{6}$Be products $\alpha$+$p$+$p$ provides complete kinematics measurement, and we can consider the population and the decay of the $^{6}$Be system in detail.
Bellow we will focus on the parameters of the model related to the reaction mechanism and how they affect on the measured spectra formation.

The data analysis is performed by comparison of experimental data with Monte-Carlo (MC) simulations  based on the three-body decay model taking into account the population of the $0^+$ and $2^+$ states only, see Section~\ref{sec:theory}. Observables relevant to the $^{6}$Be decay will be treated in a specific $^{6}$Be centre-of-mass frame with $Z$ axis directed along the transferred momentum vector.

\begin{figure}[tb]
\centering
	\begin{overpic}[width=0.9\linewidth]{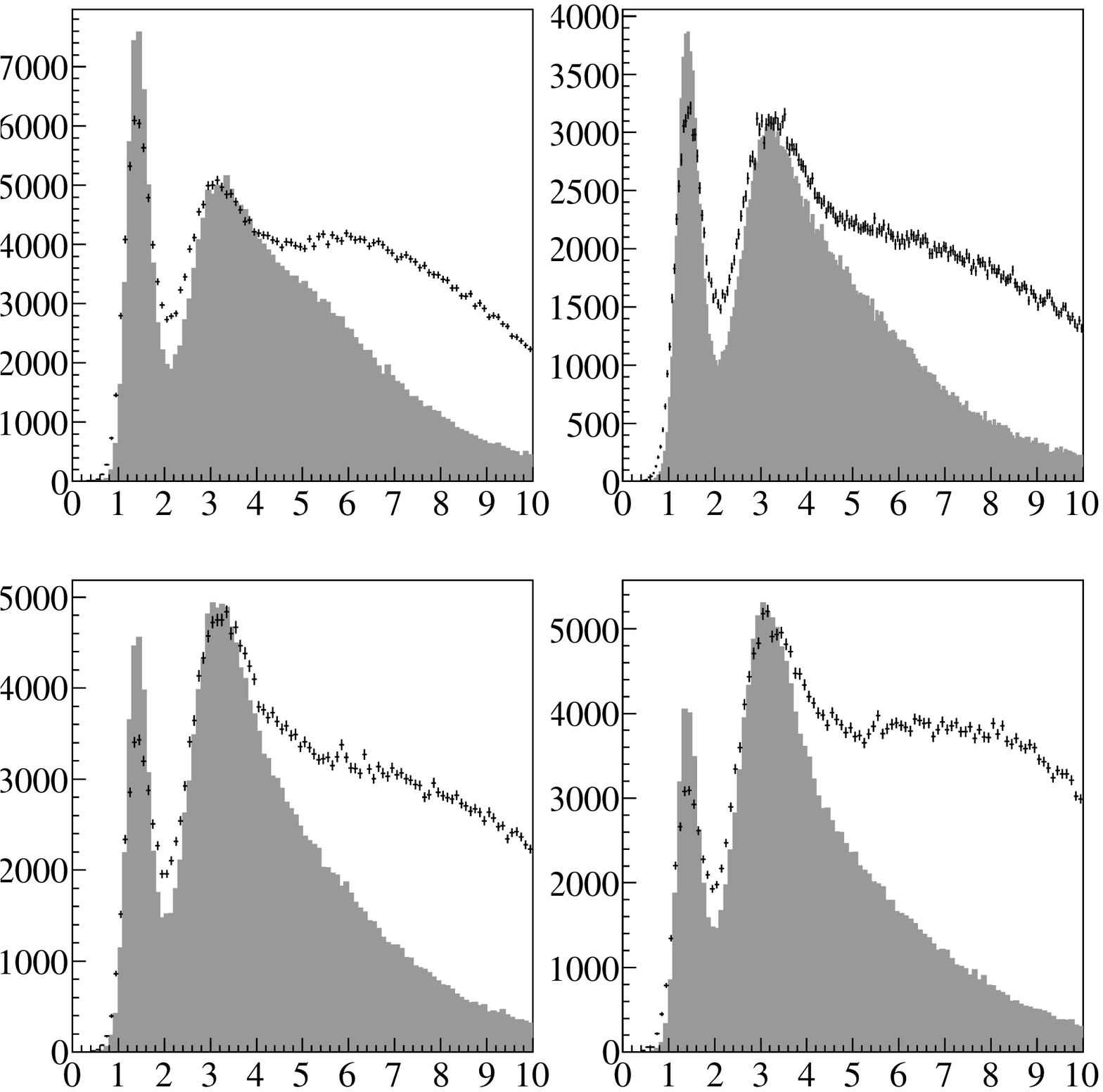}
		\put(-5,42){\rotatebox{90}{\large Events}}
		\put(20,90){\fcolorbox{white}{white}{\small \thetaBe$\in$(45,60)$^\circ$}}
		\put(68,90){\fcolorbox{white}{white}{\small \thetaBe$\in$(60,75)$^\circ$}}
		\put(23,42){\rotatebox{0}{\small \thetaBe$\in$(75,90)$^\circ$}}
		\put(70,42){\rotatebox{0}{\small \thetaBe$\in$(90,120)$^\circ$}}
	\end{overpic}
	\begin{overpic}[width=0.9\linewidth]{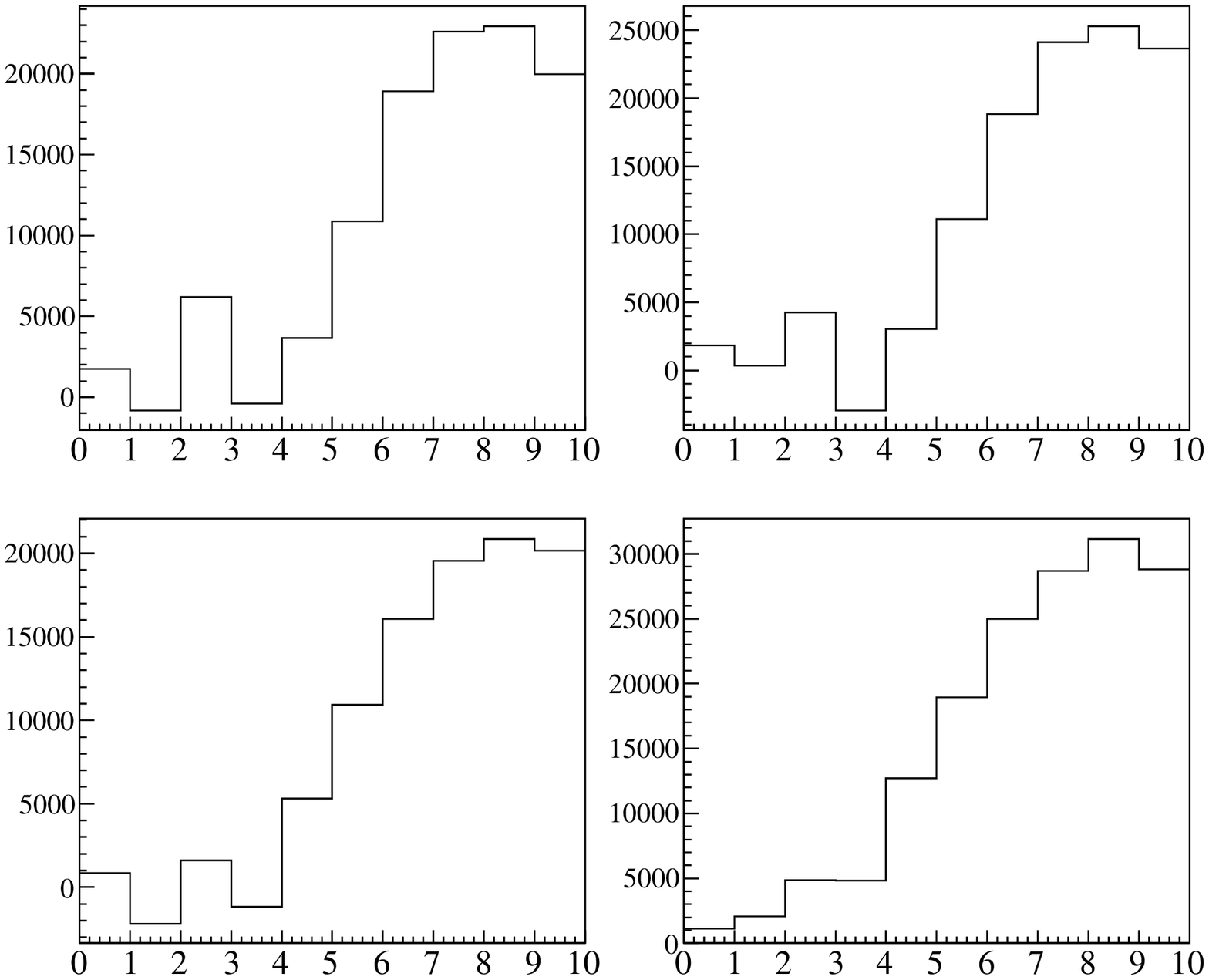}
		\put(-5,42){\rotatebox{90}{\large Events}}
		\put(9,78){\rotatebox{0}{\small \thetaBe$\in$(45,60)$^\circ$}}
		\put(59,78){\rotatebox{0}{\small \thetaBe$\in$(60,75)$^\circ$}}
		\put(9,35){\rotatebox{0}{\small \thetaBe$\in$(75,90)$^\circ$}}
		\put(59,35){\rotatebox{0}{\small \thetaBe$\in$(90,120)$^\circ$}}
		\put(41,-3){\rotatebox{0}{\large $E_T$ (MeV)}}
	\end{overpic}
		%
%
\caption{Comparison experimental data and simulations including $0^+$ and $2^+$ states in different intervals of \thetaBe. In four upper panels the experimental and simulated data are depicted by crosses and grey histogram, respectively. In four lower panels the subtraction of experimental and simulated histograms with energy bin of 1 MeV are shown for the same angular ranges. This subtraction shows the expected contributions of $J^-$ states interpreted as isovector soft dipole mode contribution.}
\label{fig:BeCorrFigSpectra}
\end{figure}

We have treated our data in the whole angular range of \thetaBe\ but we will make emphasis on the analysis of the region of $\theta_{\rm Be} \in (45, 120)^\circ$ and $E_T < 3.1$ MeV [see rectangle in Fig.~\ref{fig:IMspectrum} (b)]. Both, the ground and first excited states, are well pronounced and are measured with sufficient statistics in this region. At smaller angles setup efficiency is severely suppressed by the telescope acceptance, while at larger angles population cross section is quite low. The model calculations were passed through a virtual measuring setup taking into account all major details of the experimental setup.

We will attempt to compare theoretical results with experimental data by fitting the three aspects of the density matrix related to investigated states:

\noindent (i) Population ratio of the $0^+$ ground state to the $2^+$ first excited state;

\noindent (ii) Intensity of the spin-alignment for population of the $2^+$ state;

\noindent (iii) Interference phase between of the $0^+$ and $2^+$ states.


\subsection{Population rates for $0^+$ and $2^+$}
\label{sec:popul}


Comparison of the simulated and experimental data in different angular intervals is given in Fig.\ \ref{fig:BeCorrFigSpectra}.
Experimental and simulated data are depicted by crosses and gray histograms, respectively. We can see that MC simulations overestimates a bit the energy resolution of the experimental setup. For that reason the simulated data were fitted to experimental ones by comparing the numbers of events corresponding to the population of ground state $E_T< 2$\,MeV and those forming the left slope of the $2^+$ state peak $2.5< E_T< 3.1$\,MeV.

In contrast with treatment of Ref.\ \cite{Fomichev:2012}, here we do not include in the MC simulations the contribution of the $J^-$ continuum (Isovector Soft Dipole Mode contribution). Instead, in lower panels of Fig.\ \ref{fig:BeCorrFigSpectra} we show results of the subtraction of simulated $0^+$ and $2^+$ contributions from the experimental spectrum. We can see that the IVSDM contributions are weakly dependent on the angular range. Another important thing we realize from this illustration is a significant contribution of the IVSDM for the right wing of the $2^+$ resonance. This message is confirmed by theoretical calculations of Ref.\ \cite{Fomichev:2012}, also shown in Fig.\ \ref{fig:IMspectrum} (c). We see that if we would like to study the $0^+/2^+$ mixing only we should stick our analysis mainly to the left wing of the $2^+$ resonance with $E_T< 3.1$\,MeV.


\subsection{Spin-parity identification and density matrix parametrization}
\label{sec:densityMatrix}


Before we turn to charge exchange-reaction, some explanation how spin-parity identification based on density-matrix formalism was realized in our previous works is needed. The $(t,p)$ reactions were used for population of three-body continuum states in $^{5}$H and $^{10}$He in Refs.~\cite{Golovkov:2004,Golovkov:2005,Sidorchuk:2012}. Such two-neutron transfer reactions seem to be reliably described by ``dineutron'' transfer (two nucleons are transfered in a state with $S$=0).
Then suppositions about Eq.\ (\ref{eq:alignment}) are fully valid and we get highly aligned density matrix in the transfered momentum frame with practically complete ``polar'' alignment
\begin{equation}
\rho^{J'M'}_{JM} \sim \delta_{M, \pm 1/2}\,\delta_{M', \pm 1/2} \quad \mbox{or}
\quad \rho^{J'M'}_{JM} \sim \delta_{M,0}\,\delta_{M',0}\,,
\label{eq:polar}
\end{equation}
for half-integer and integer spin of the initial system, respectively. Such strong alignment guarantee very expressed interference patterns for broad overlapping continuum states which were used in the data analysis \cite{Golovkov:2005,Sidorchuk:2012}.

In general case the angular distribution of two-body decays is expressed in terms of associated Legendre polynomials $P^M_L(x)$.
If a polar-aligned state with angular momentum $L$ decays via emission of particle with $J$=0 the angular distribution of the products may be expressed as
\begin{equation}
\frac{d \sigma}{d \cos \theta} \sim |P_L^0(\cos \theta)|^2 \,,
\label{eq:legendre}
\end{equation}
for the selected alignment system, producing expressed and easy-to-interpret angular distribution.

In the case of the three-body decay, there exist an evident limit effectively reducing three-body motion to two-body motion:
\begin{equation}
\varepsilon \rightarrow 0 \,.
\label{eq:3bodyTo2body}
\end{equation}
When the relative motion of one pair of particles (e.g.\ two protons) is fully suppressed and the three-body decay is determined as two-body motion of alpha-particle and diproton with zero energy.
In this limit we are getting for three-body decays the same very expressed angular distributions, but now in the corresponding $\beta$-angle [see Fig.~\ref{fig:scheme}(c)].

We introduce the term \emph{quasibinary kinematic} for three-body decay when the condition (\ref{eq:3bodyTo2body}) is replaced by $\varepsilon<x$ assuming a choice made for the upper limit of $\varepsilon$ providing satisfactory accuracy for the studied process. For high-statistics measurements the value $x$ can be gradually reduced to reveal expressed and easy-to-interpret correlation patterns.

In our analysis we fix some $E_T$ and \thetaBe\ ranges and consider different correlation patterns within them. For internal correlations we consider $\varepsilon$ and $\theta_k$ distributions and for external correlations the most interesting are angular distributions $\theta_{\alpha}$ of $\alpha$-particles in the momentum transfer system, $\theta_{\alpha}=\pi-\beta$.

Formally, the terms of the density matrix $\rho^{J'M'}_{JM}(q,E_T)$ for the pole reaction mechanism in Eq.\ (\ref{eq:rho-matr}) depend on two parameters: $q$ and $E_T$. Looking in Fig.\ \ref{fig:IMspectrum} (d) it is easy to find that for energy and angular range of our interest momentum transfer depends only on angle, not on energy and thus it is reliable to consider the $\rho^{J'M'}_{JM}$ dependence on $E_T$ and $\theta_{\text{Be}}$ in a factorized form. So, we presume that:

\noindent (i) The energy profile of the $0^+$ and $2^+$ states individually is defined by the energy dependence of the three-body amplitudes $A_{JMSM_S}(E_T,\Omega_{\varkappa})$ as provided by three-body theoretical calculations.

\noindent (ii) The ``global'' population rate for the $0^+$ and $2^+$ states as fitted to experiment is defined by the parameter $\rho^{00}_{00}/\sum_M \rho^{2M}_{2M}$.

\noindent (iii) The following items are considered separately for each $\{E_T,\theta_{\text{Be}} \}$ bin: the alignment for the $2^+$ state ($\rho^{2M}_{2M}$ dependence on $M$) and the ``interference angle'' $\varphi_{02}$  between $0^+$ and $2^+$ states, which define off-diagonal density matrix term parameterised as
\begin{equation}
\rho^{00}_{20} = \rho^{20}_{00} =\sqrt{\rho^{00}_{00} \, \rho^{20}_{20}} \, \cos (\varphi_{02})\,.
\label{eq:rho-off-diag}
\end{equation}
%

%
%

\begin{figure}[tb]
	\begin{center}
		\includegraphics[width=0.245\textwidth]{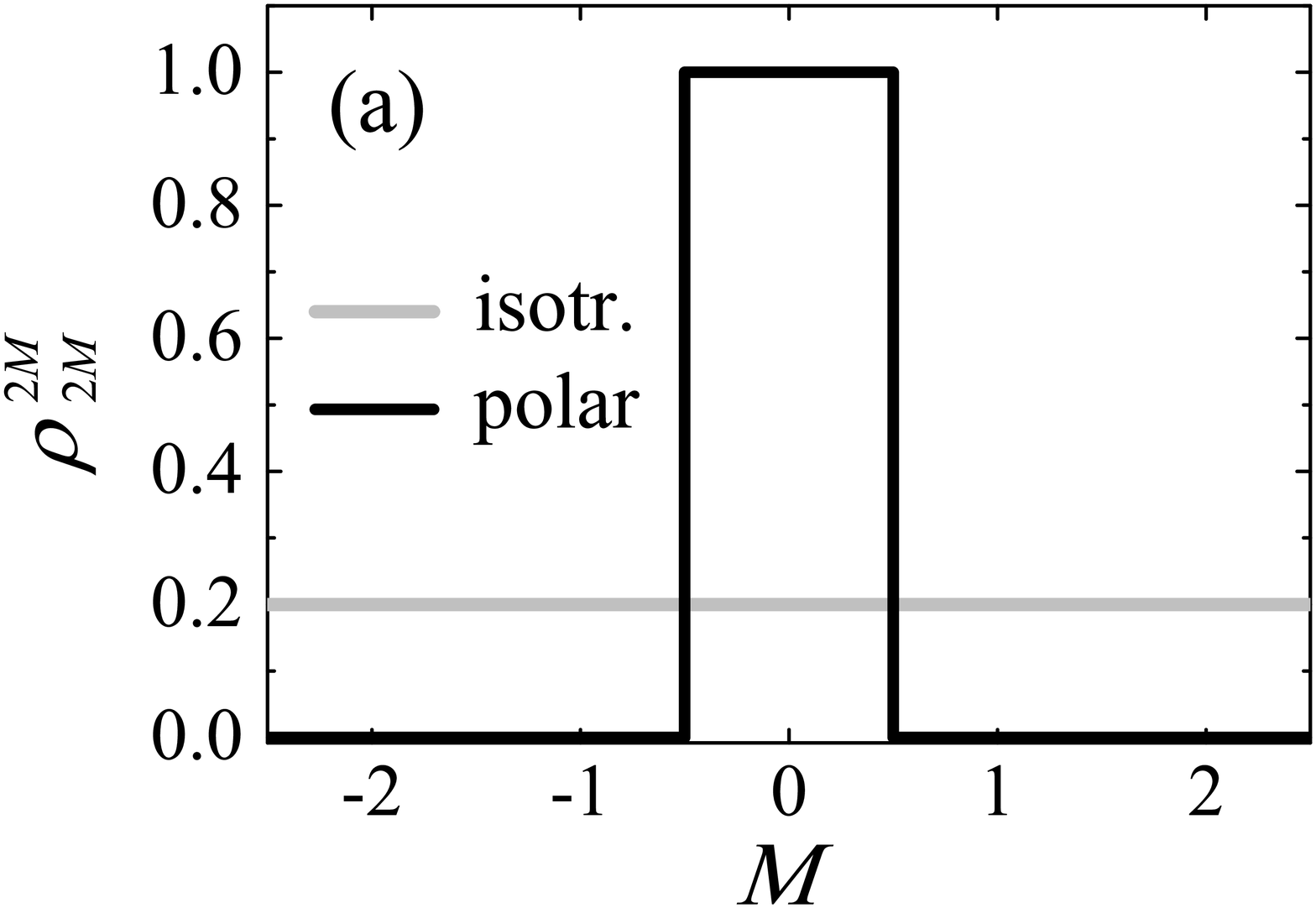}
		\includegraphics[width=0.224\textwidth]{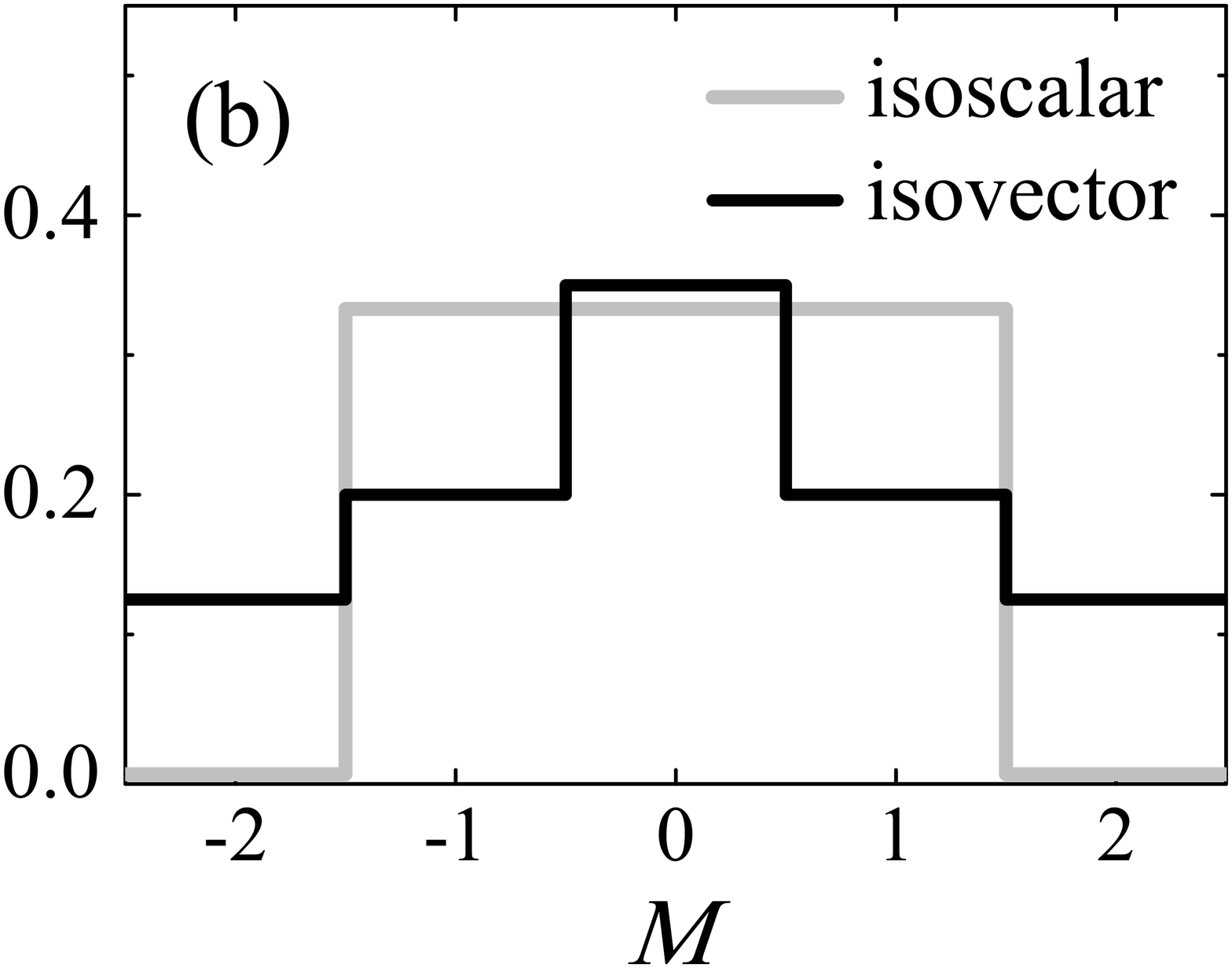}
	\end{center}
	\caption{The density matrix (\ref{eq:rho-matr}) spin structure for the $2^+$ state. (a) Model approximations of isotropic and polar alignments actually used for MC simulations. (b) Isovector-scalar (constant $c_{ivs}$) and isovector-vector (constant $c_{ivv}$) couplings for potential (\ref{eq:wig-pot}).}
	\label{fig:rho-matr}
\end{figure}

The expected alignment pattern for the $^{6}$Be $2^+$ state populated in the charge-exchange reaction induced by the potential Eq.\ (\ref{eq:wig-pot}) is illusrated in Fig.\ \ref{fig:rho-matr} (a). Actually, more expressed alignment parameterizations were used for the MC simulations. Expressions
\begin{eqnarray}
\label{eq:RhoAl}
\rho_{2M} \sim \delta_{M0} , \\
\label{eq:RhoNoAl}
\rho_{2M} \sim 1/5  ,
\end{eqnarray}
correspond to population of fully polar-aligned \eqref{eq:RhoAl} and nonaligned (isotropic) \eqref{eq:RhoNoAl} $2^+$ state, respectively. If we consider the angular distribution of $\alpha$-particle fragment in the momentum transfer frame, then the isotropic density matrix should provide isotropic angular distribution for the isolated $2^+$ state. Note that in the case of significant interference with other states an anisotropic distribution can be obtained even for isotropically populated state. With increase of the alignment more and more distinctive form of angular distribution should be obtained for the $2^+$  state, tending to $|P_2^0(\cos \theta_{\alpha})|^2$ in the limit of polar alignment and the under the condition \eqref{eq:3bodyTo2body}.

Three extreme cases of interference between $0^+/2^+$, described by angle \gPhase, are considered. We simulated the constructive interference (\gPhase$ =0^\circ$), destructive interference (\gPhase$ =180^\circ$) and situation when amplitudes of $0^+$ and $2^+$ states are summed incoherently (\gPhase$=90^\circ$) for both cases determined by equations \eqref{eq:RhoAl} and \eqref{eq:RhoNoAl}.

So, within this paper, six special cases are systematically illustrated, those given by two extreme cases of $2^+$ alignment and three distinct cases of $0^+/2^+$ interference, see e.g.\ Fig.\ \ref{fig:GSepsilonT}.


\subsection{Ground state correlations}


We start our analysis from the part of the $^{6}$Be excitation spectrum where the $0^+$ ground state is only present. To eliminate the possible effects caused by the interference with the $2^+$ state we restricted analysis here to excitation energy $E_T$$\leq$1.4\,MeV, where contribution of the left ``wing'' of the first excited state can be reliably neglected.
So, we have no free model parameters related to the reaction mechanism ($0^+$ state by itself is ``isotropic'' by definition) there and internal correlations of $^{6}$Be decay products should be the same for the whole range of angle \thetaBe.

\begin{figure}[!ht]
	\centering
	\begin{overpic}[width=0.99\linewidth]{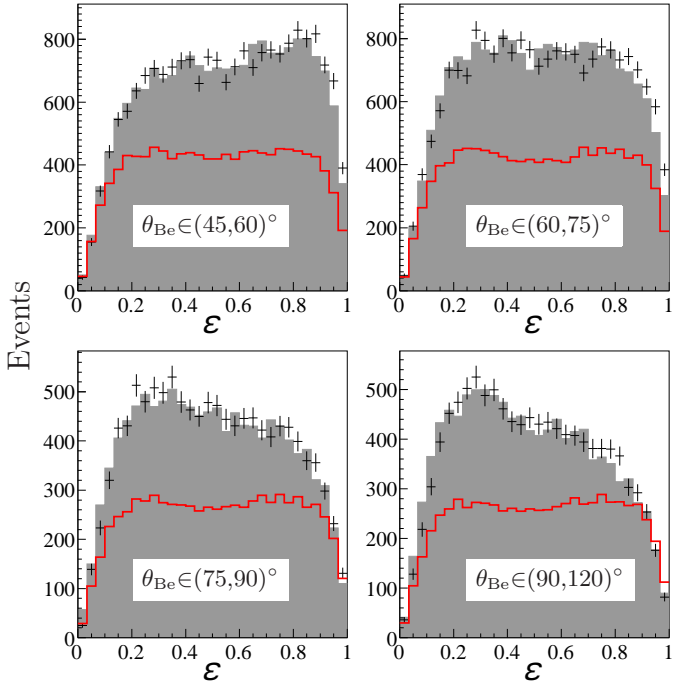}
		\put(-4,45){\rotatebox{90}{\large Events}}
		\put(14,65){\fcolorbox{white}{white}{\small \thetaBe$\in$(45,60)$^\circ$}}
		\put(62,65){\fcolorbox{white}{white}{\small \thetaBe$\in$(60,75)$^\circ$}}
		\put(14,14){\fcolorbox{white}{white}{\small \thetaBe$\in$(75,90)$^\circ$}}
		\put(62,14){\fcolorbox{white}{white}{\small \thetaBe$\in$(90,120)$^\circ$}}
	\end{overpic}
	\caption{Energy distributions $\varepsilon_T$ for $^{6}$Be decay with $E_T$$<$1.4\,MeV for different \thetaBe\ bins. Experimental and simulated data are depicted by crosses and grey histograms, respectively. Theoretical input is illustrated by red lines.}
	\label{fig:GSepsilonT}
\end{figure}

We can see in Fig.~\ref{fig:GSepsilonT} that observed $\varepsilon_T$ distributions are qualitatively different for different \thetaBe\ bins.
In spite of this fact the simulated energy distributions (gray histograms) are in a nice agreement with experimental data (theoretical distributions depicted by red histograms are the same for all panels in Fig.\ \ref{fig:GSepsilonT}).

\begin{figure}[tb]
	\centering
	\begin{overpic}[width=0.99\linewidth]{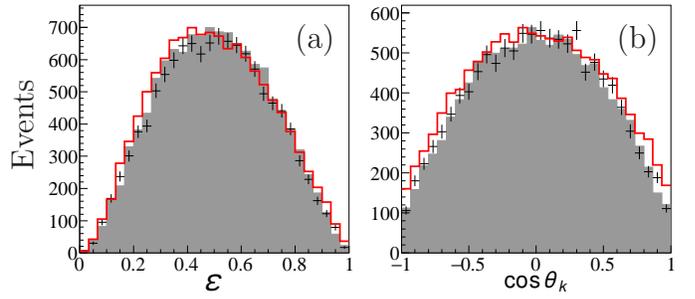}
		\put(-4,19){\rotatebox{90}{\large Events}}
		\put(40,38){\rotatebox{0}{\large (a)}}
		\put(90,38){\rotatebox{0}{\large (b)}}
	\end{overpic}
	\caption{Panel (a) shows energy distributions $\varepsilon_Y$ in the Jacobi ``Y''-system. Panel (b) shows angular distribution $\cos \theta_k^{(T)}$ in the Jacobi ``T''-system. Kinematical range $E_T$$<$1.4\,MeV and 75$^\circ$$<$$\theta_{\text{\scriptsize Be}}$$<$90$^\circ$.}
	\label{fig:GSotherCorrs}
\end{figure}

The effect of the response of the experimental setup is much smaller for $\varepsilon_Y$ distribution in the ``Y''-system and $\cos \theta_k^{(T)}$  in the ``T''-system. It also only very weakly depending on the kinematical range of \thetaBe.
In Fig.~\ref{fig:GSotherCorrs} we show typical picture of these distributions. 
Corrections induced by detection efficiency are noticeable, but not large.

Analysis of internal correlations for the ground state given here may be seen as a benchmark in two ways.
On the one hand, it provides a confirmation of the theoretically predicted correlations, which were already tested against highly detailed experimental data of works~\cite{Grigorenko:2009c,Egorova:2012}.
Thus, the full consistency of our experiment with previous high-precision experiments \cite{Grigorenko:2009c,Egorova:2012} is demonstrated.
On the other hand, the nice agreement in Figs.~\ref{fig:GSepsilonT} and \ref{fig:GSotherCorrs} means that the MC simulation is working reliably in the whole considered \thetaBe\ range and well represents response of the experimental setup.
This is an important prerequisite for the next more complicated steps of our analysis.


\subsection{Correlations at the right slope of the $0^+$ state}


Effects of $0^+/2^+$ interference become important already on the right slope of the ground $0^+$ state. Let us have a look at the energy range 1.4$<$$E_T$$<$1.9\,MeV. If we look in theoretical predictions shown in Fig.\ \ref{fig:IMspectrum} (c), we can find that the relative probability of the $2^+$ state population expected in this energy range is just around $1\%$ of the $0^+$ one.
Nevertheless, it is sufficient to produce a significant modification in the correlations. This is the important motivation for use of correlations as a tool for studies: they are sensitive to amplitudes, not to probabilities.
Therefore, the effects of even small-weight configurations can be drastically amplified.

\begin{figure}[tb]
	\centering
	\begin{overpic}[width=0.99\linewidth]{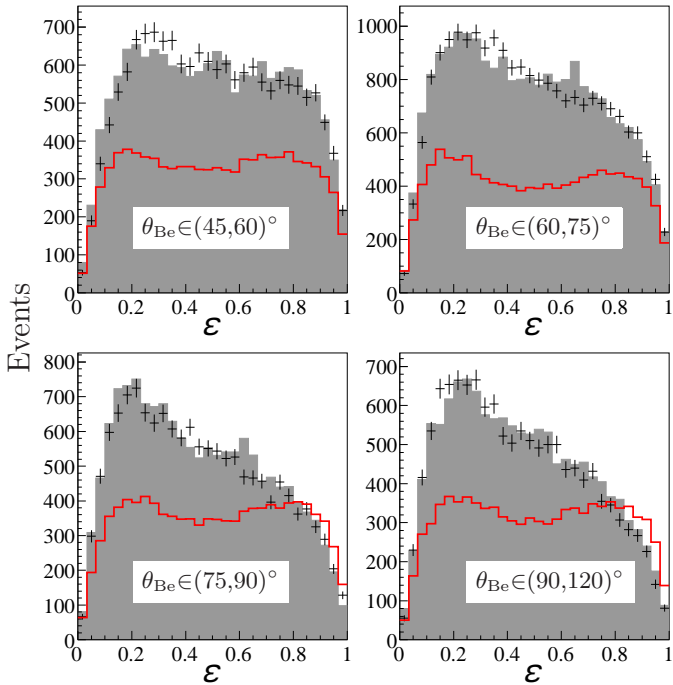}
		\put(-4,45){\rotatebox{90}{\large Events}}
		\put(14,65){\fcolorbox{white}{white}{\small \thetaBe$\in$(45,60)$^\circ$}}
		\put(62,65){\fcolorbox{white}{white}{\small \thetaBe$\in$(60,75)$^\circ$}}
		\put(14,14){\fcolorbox{white}{white}{\small \thetaBe$\in$(75,90)$^\circ$}}
		\put(62,14){\fcolorbox{white}{white}{\small \thetaBe$\in$(90,120)$^\circ$}}
	\end{overpic}
	\caption{Energy distribution $\varepsilon_T$ for $^{6}$Be decay with 1.4$<$\Et$<$1.9\,MeV (right slope of $0^+$) for different \thetaBe\ bins. The simulation model settings are isotropic $2^+$ state (NA) and no $0^+/2^+$ interference ($\varphi_{02}=90^\circ$).}
	\label{fig:BeCorrEpsilonEvolution}
\end{figure}

First, we consider the evolution of energy $\varepsilon_T$ distributions with angle $\theta_{\text{\scriptsize Be}}$.
It is illustrated in Fig.\ \ref{fig:BeCorrEpsilonEvolution} for the ``trivial'' case of isotropic and not interfering $2^+$ state.
The calculated distributions are not much different from the ones shown in Fig.\ \ref{fig:GSepsilonT} and evidently do not depend on the angle $\theta_{\text{\scriptsize Be}}$.
However, observable $\varepsilon_T$ distributions are strongly sensitive to the angle $\theta_{\text{\scriptsize Be}}$ and we can see that MC simulations are reliably taking the experimental efficiency into account in this range as well.

\begin{figure}[tb]
	\centering
	\begin{overpic}[width=0.99\linewidth]{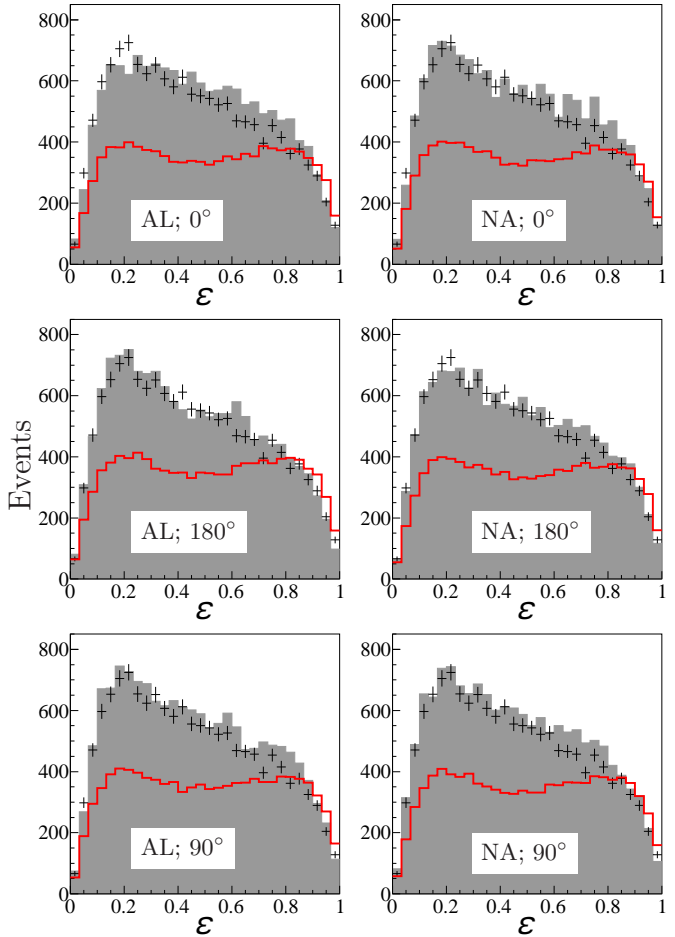}
		\put(-2,45){\rotatebox{90}{\large Events}}
		\put(11,75){\fcolorbox{white}{white}{\small AL; 0$^\circ$}}
		\put(47,75){\fcolorbox{white}{white}{\small NA; 0$^\circ$}}
		\put(11,42){\fcolorbox{white}{white}{\small AL; 180$^\circ$}}
		\put(47,42){\fcolorbox{white}{white}{\small NA; 180$^\circ$}}
		\put(11,8.5){\fcolorbox{white}{white}{\small AL; 90$^\circ$}}
		\put(47,8){\fcolorbox{white}{white}{\small NA; 90$^\circ$}}
	\end{overpic}
	\caption{Energy distributions $\varepsilon_T$ for $^{6}$Be decay with 1.4$<$\Et$<$1.9\,MeV (right slope of $0^+$) for $75^\circ$$<$$ \theta_{\text{\scriptsize Be}}$$<$90$^\circ$ and different alignment/interference settings. The left column, compares data with  theoretical model describing fully polar-aligned (AL) $2^+$ state. The right column, corresponds to the isotropic (nonaligned, NA) $2^+$ state. Upper, middle, and bottom rows correspond to interference phase $\varphi_{02}$ equal to  $0^\circ$, $180^\circ$, and $90^\circ$, respectively.}
	\label{fig:BeCorrEpsilon7590}
\end{figure}

Much more fine effect of the alignment/interference on the energy distribution $\varepsilon_T$ is illustrated in Fig.\ \ref{fig:BeCorrEpsilon7590}. We can see in this plot that there is weak dependence of the \emph{observed} shape of the distribution on the alignment/interference settings. Please note that from theoretical point of view there is no dependence of the $\varepsilon_T$ distributions on the reaction mechanism.
However, such a sensitivity of \emph{observable} distributions arise in the experimental conditions, when isotropic efficiency for registration of decay fragments in not available.
This effect has been already pointed in Ref.\ \cite{Grigorenko:2013} (see Fig.\ 6 of this work) for the $^{6}$Be $2^+$ data from experiment \cite{Egorova:2012}.

\begin{figure}[tb]
	\centering
	\begin{overpic}[width=0.99\linewidth]{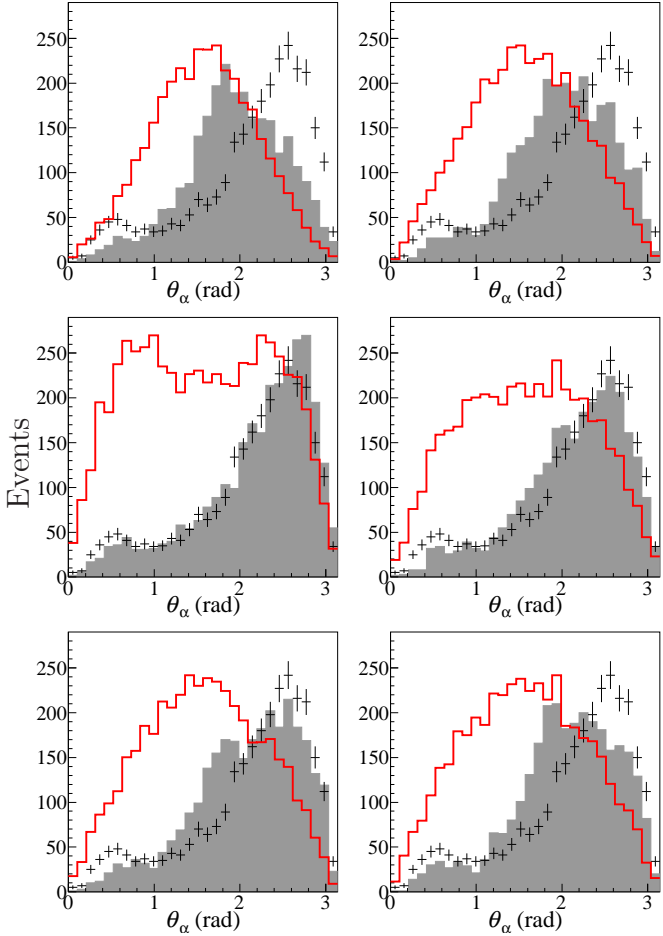}
		\put(-2,45){\rotatebox{90}{\large Events}}
	\end{overpic}
	\caption{Angular distributions for the $\alpha$-particle emission in the momentum transfer frame in the range 1.4$<$\Et$<$1.9\,MeV and 75$^\circ$$<$$ \theta_{\text{\scriptsize Be}}$$<$90$^\circ$. Alignment/interference settings are the same as in Fig.~\ref{fig:BeCorrEpsilon7590}.}
	\label{fig:GSThetaAright7590}
\end{figure}

The dependence of Fig.\ \ref{fig:BeCorrEpsilon7590} is quite curious, but too weak for practical application and deriving definite conclusions.
To distinguish clearly the effects of alignment/interference it is better to consider external correlations in the momentum transfer frame.
Angular distributions for $\alpha$-particle emission in the momentum transfer frame are illustrated in Fig.\ \ref{fig:GSThetaAright7590}.
We analysed the $^{6}$Be decay in quasibinary approximation under the condition
\begin{equation}\label{eq:epsilonCondBe}
	\varepsilon_T < 0.2,
\end{equation}
which ensured high enough statistics for the considered $\{E_T,\theta_{\text{\scriptsize Be}} \}$  windows.
It can be seen in Fig.\ \ref{fig:GSThetaAright7590} that already theoretical angular distributions are very sensitive to alignment/interference conditions. 
This sensitivity is further enhanced by imperfect experimental efficiency. It is clear that these distributions can be used to fix alignment/interference parameters with reasonable confidence.
The analysis analogous to that of Fig.\ \ref{fig:GSThetaAright7590} was performed in the whole $\theta_{\text{\scriptsize Be}}$ range and the results are summarized in Table \ref{tab:results}.

Note, that such a strong sensitivity of the observed angular distributions is obtained just for $\sim$1\% of the $2^+$ state relative weight in the considered $E_T$ energy window.


\subsection{Correlations at the left slope of the $2^+$ state}


We may expect that effects of alignment/interference will be more pronounced in the region with higher probability of population of the $2^+$ state.
As illustration we provide here some details for the range 2.5$<$\Et$<$3.1\,MeV. This corresponds to the left (rising) slope of the first excited $2^+$ state of $^{6}$Be. It can be expected from Fig.\ \ref{fig:IMspectrum} (c) $\sim 20 \%$ of $0^+$ contribution in this range making strong interference highly probable.  Certain ``contamination'' of the correlations in this range by $J^-$ contributions can be expected, but analysis shows that in reality it appears to be not of importance.
Thus, the analysis scheme here is quite stereotypical with that of the previous Section.

\begin{figure}[tb]
	\centering
	\begin{overpic}[width=0.9\linewidth]{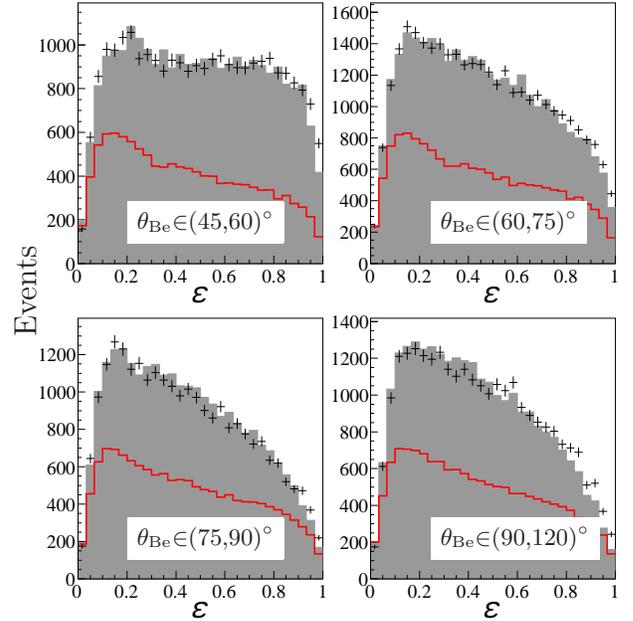}
		\put(-4,45){\rotatebox{90}{\large Events}}
		\put(14,62){\fcolorbox{white}{white}{\small \thetaBe$\in$(45,60)$^\circ$}}
		\put(62,62){\fcolorbox{white}{white}{\small \thetaBe$\in$(60,75)$^\circ$}}
		\put(14,12){\fcolorbox{white}{white}{\small \thetaBe$\in$(75,90)$^\circ$}}
		\put(62,12){\fcolorbox{white}{white}{\small \thetaBe$\in$(90,120)$^\circ$}}
	\end{overpic}
	\caption{Energy distribution $\varepsilon_T$ for the $^{6}$Be decay with 2.5$<$\Et$<$3.1\,MeV (left slope of $2^+$) for different \thetaBe\ bins. The simulation model settings are isotropic $2^+$ state (NA) and no $0^+/2^+$ interference ($\varphi_{02}=90^\circ$).}
	\label{fig:ESepsilonTleft}
\end{figure}

\begin{figure}[tb]
	\centering
	\begin{overpic}[width=\linewidth]{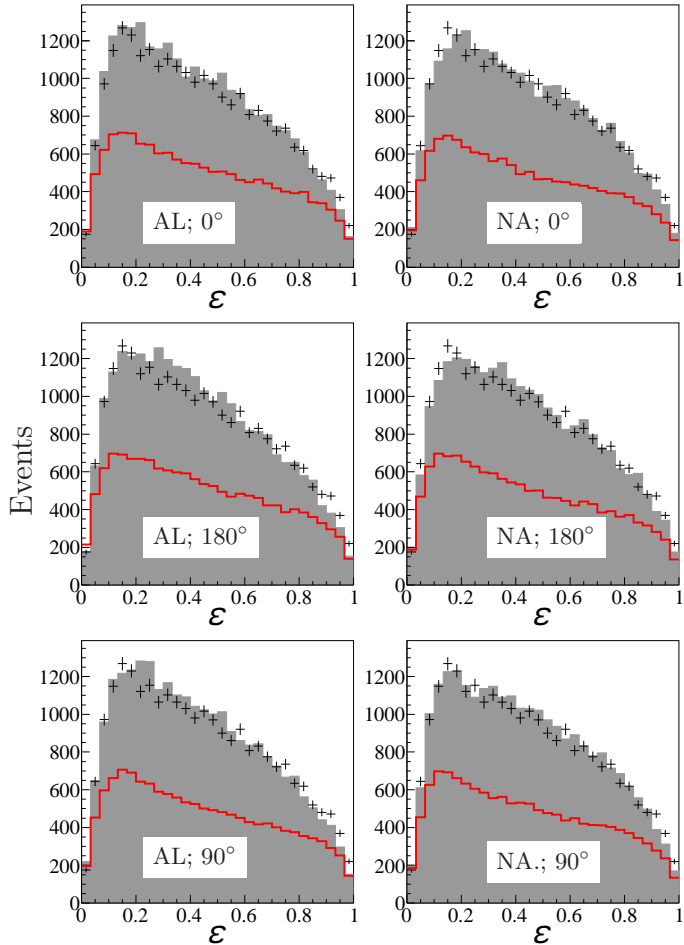}
		\put(-3,45){\rotatebox{90}{\large Events}}
		\put(11,75){\fcolorbox{white}{white}{\small AL; 0$^\circ$}}
		\put(47,75){\fcolorbox{white}{white}{\small NA; 0$^\circ$}}
		\put(11,42){\fcolorbox{white}{white}{\small AL; 180$^\circ$}}
		\put(47,42){\fcolorbox{white}{white}{\small NA; 180$^\circ$}}
		\put(11,8.5){\fcolorbox{white}{white}{\small AL; 90$^\circ$}}
		\put(47,8){\fcolorbox{white}{white}{\small NA.; 90$^\circ$}}
	\end{overpic}
	\caption{Energy distribution $\varepsilon_T$ for $^{6}$Be decay with 2.5$<$\Et$<$3.1\,MeV (left slope of $2^+$) for $75^\circ$$<$$ \theta_{\text{\scriptsize Be}}$$<$$90^\circ$ and different alignment/interference settings, see also caption of the Fig.~\ref{fig:BeCorrEpsilon7590} for details.}
	\label{fig:ESepsilonT7590left}
\end{figure}

Our first test is energy distribution $\varepsilon_T$, which gives minimal validation of the MC procedure quality, see Fig.\ \ref{fig:ESepsilonTleft}.
This energy distribution for the $2^+$ state is qualitatively different from that for the $0^+$ state.
The major effects of experimental response are effectively removed by MC simulations shown in Fig.\ \ref{fig:ESepsilonTleft}.
More fine effects of the alignment/interference on the observable distributions are illustrated for the selected $\theta_{\text{\scriptsize Be}}$ range in Fig.\ \ref{fig:ESepsilonT7590left}.

All other distributions related to internal correlations, \thetaK\ in both ``Y'' and ``T'' systems and $\varepsilon_Y$ show the same nice agreement between experiment and theory. As a result of these studies we can declare two observations:

\noindent (i) The internal correlations do not seem to demonstrate noticeable dependence on the population conditions. This is a quite expected result for the narrow $0^+$ state ($\Gamma \approx 90$\,keV), however, for much broader $2^+$ state ($\Gamma \sim 1$\,MeV) this is not evident in advance.
Thus, the internal motion of the three-body system seems to be really disentangled from the motion of the three-body system as a whole as it is presumed in the density matrix formalism.

\noindent (ii) The same theoretical input for the $2^+$ state correlations was used for MC simulations in Ref.\ \cite{Egorova:2012}.
As far as the agreement between theory and experiment was also very good in this work, it means that there is a complete agreement between this experiment and the experiment  \cite{Egorova:2012}.
The $^{6}$Be states were populated in a high-energy ($E_{\rm beam} \sim 70$\,AMeV) knockout from $^{7}$Be beam in experiment \cite{Egorova:2012}.
This is very different reaction mechanism, so the internal correlations in the decay of relatively broad $2^+$ state seem to be not sensitive also to this aspect of the reaction mechanism.

\begin{figure}[tb]
	\centering
	\begin{overpic}[width=\linewidth]{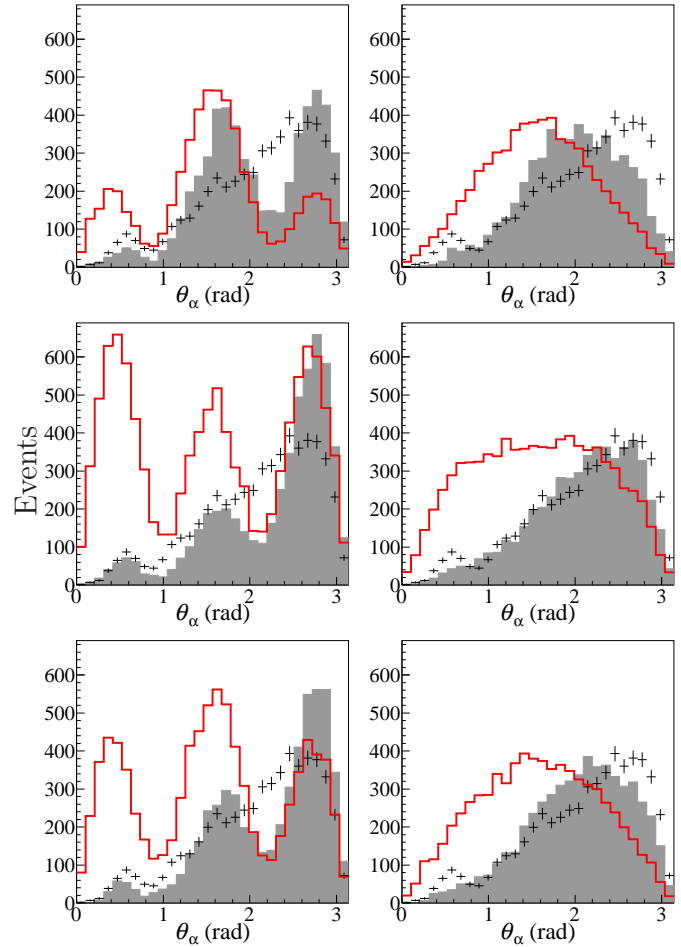}
		\put(-2,45){\rotatebox{90}{\large Events}}
	\end{overpic}
	\caption{Angular distributions for the $\alpha$-particle emission in the momentum transfer frame in the range 2.5$<$\Et$<$3.1\,MeV and $75^\circ$$<$$\theta_{\text{\scriptsize Be}}$$<$$90^\circ$. Alignment/interference settings are the same as in Fig.~\ref{fig:BeCorrEpsilon7590}.}
	\label{fig:ESThetaAleft7590}
\end{figure}

\begin{figure}[tb]
	\centering
	\begin{overpic}[width=\linewidth]{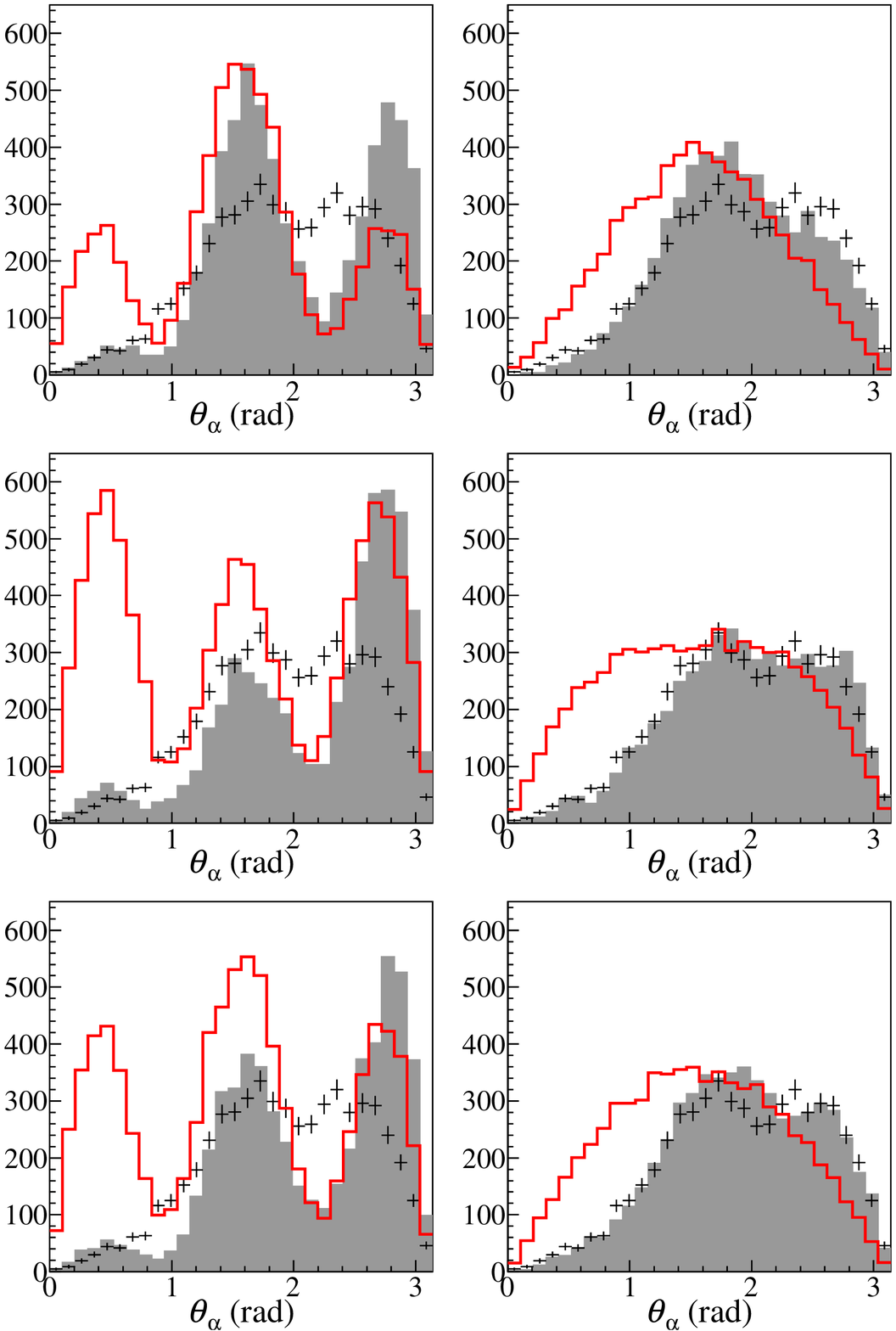}
		\put(-2,45){\rotatebox{90}{\large Events}}
	\end{overpic}
	\caption{Angular distributions for the $\alpha$-particle emission in the momentum transfer frame in the range 2.5$<$\Et$<$3.1\,MeV and $90^\circ$$<$$\theta_{\text{\scriptsize Be}}$$<$$120^\circ$. Alignment/interference settings are the same as in Fig.~\ref{fig:BeCorrEpsilon7590}.}
	\label{fig:ESThetaAleft90120}
\end{figure}

The analysis of external correlations is illustrated by Figs.\ \ref{fig:ESThetaAleft7590} and \ref{fig:ESThetaAleft90120} for two selected $\theta_{\text{\scriptsize Be}} $ ranges. Again the sensitivity of the $\theta_{\alpha}$ angular distributions to the alignment/interference conditions is very high. However, we can find the density matrix parameters for which near perfect description of the distribution is provided. The results of our fits are summarized in the Table \ref{tab:results}. Part of the spectrum characterized by pure ground state (\Et$<$1\,MeV) does not depend on the density matrix parameters and it is not shown in the Table.

\begin{table*}[tb]
\caption{The best fit to experimental data of density matrix parameters for different $\{E_T,\theta_{\text{Be}} \}$ ranges. The fits were found using the figures with \thetaAlpha\ distribution for all six configurations of the theoretical model.}
\begin{ruledtabular}
\begin{tabular}[c]{ccccc}
$E_T$ (MeV)  & \thetaBe$\in$(45,60)$^\circ$ & \thetaBe$\in$(60,75)$^\circ$ & \thetaBe$\in$(75,90)$^\circ$ & \thetaBe$\in$(90,120)$^\circ$ \\
\hline
1.4--1.9  & AL; \gPhase=135$^\circ$ & AL + $50\%$ NA; \gPhase=180$^\circ$ & AL;  \gPhase=180$^\circ$ & AL + $20\%$ NA; \gPhase=180$^\circ$ \\
1.9--2.5	& AL + $50\%$ NA; \gPhase=135$^\circ$ & NA + $10\%$ AL; \gPhase=180$^\circ$ & NA; \gPhase=180$^\circ$ & AL + $10\%$ NA; \gPhase=90$^\circ$ \\
2.5--3.1  & NA + $10\%$ AL; \gPhase=180$^\circ$ & AL + $10\%$ NA; \gPhase=180$^\circ$ & NA + $30\%$ AL; \gPhase=90$^\circ$ & NA; \gPhase=135$^\circ$ \\
\end{tabular}
\end{ruledtabular}
\label{tab:results}
\end{table*}


\subsection{Correlations at the right slope of the $2^+$ state}


Let us consider now the energy range 3.1$<$\Et$<$3.7\,MeV. It has been discussed above in the Section \ref{sec:popul} that important contribution IVSDM is expected here. Inclusive contribution of $J^-$ states here can be theoretically evaluated as $\sim 25 \%$ from Fig.\ \ref{fig:IMspectrum} (c). How this fact is reflected in the correlations?

\begin{figure}[tb]
	\centering
	\begin{overpic}[width=\linewidth]{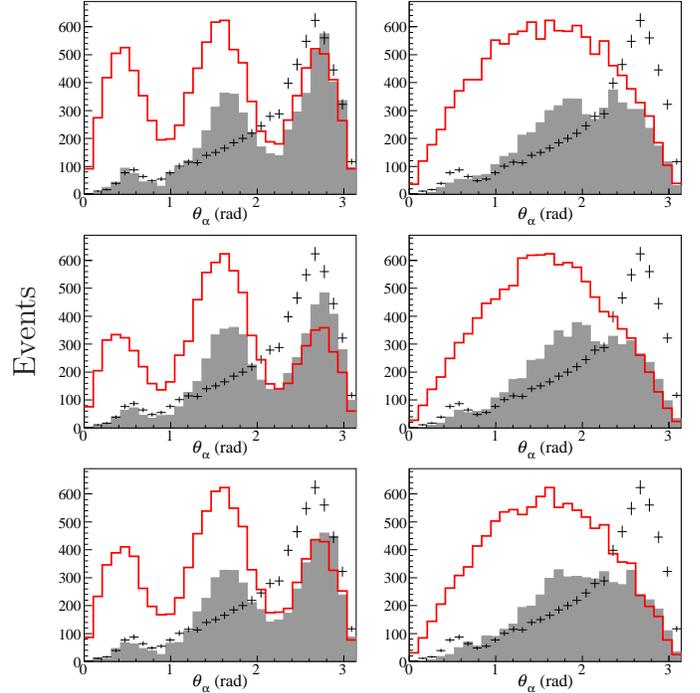}
		\put(-4,45){\rotatebox{90}{\large Events}}
	\end{overpic}
	\caption{Angular distribution of $\alpha$-particles in the momentum transfer frame for the range 3.1$<$\Et$<$3.7\,MeV and $75^\circ$$<$$\theta_{\text{\scriptsize Be}}$$<$$90^\circ$. Alignment and interference conditions are the same as in Fig.~\ref{fig:BeCorrEpsilon7590}.}
	\label{fig:ESThetaAright7590}
\end{figure}

The typical picture of comparison of theoretical data with experimental ones for energy above the peak corresponding to the $2^+$ state is shown in Fig.~\ref{fig:ESThetaAright7590}. It is obvious that experimental data cannot be fitted using $0^+$ and $2^+$ contributions only because of the simulated events excess for all model interference/alignment settings at \thetaAlpha $\sim \pi/2$. Moreover, it is clear that forward/backward asymmetry in the data is much higher than in the simulations. Such a forward/backward asymmetry is not possible for isolated states or for interference of states with the same parity. 
This means that asymmetry obtained in the simulations can be related only to the response of the experimental setup.
Simulations show that this effect is not sufficient to explain the observed forward/backward asymmetry.
It means that additional interference of $0^+$ and $2^+$ with some $J^-$ states is needed for explanation of the data. This can be seen as additional independent proof of the IVSDM contribution at $E_T > 3$ MeV.

\section{Discussion}


Studies of the three-body correlations for decays  \cite{Miernik:2007b,Grigorenko:2009,Ascher:2011} or particle emission from  states populated in reactions \cite{Golovkov:2004a,Golovkov:2005,Sidorchuk:2012,Brown:2014,Brown:2015a} are quite active in the recent years.
In such studies the experimental question arises, which should be resolved to make theoretical interpretation possible:  how much the \emph{observed} correlation patterns are different from the \emph{actual}? In this work we provide extensive illustration of this issue: even for the $^{6}$Be ground state case, the observed three-body correlation patterns demonstrate strong variation 
depending on the specific region of the kinematical space.
The influence of the experimental efficiency is especially harmful for studies of the external correlations.
In this work we disentangle the effects related to response of the experimental setup from the effects of alignment/interference for energy range where $0^+$ and $2^+$ states of $^{6}$Be effectively overlap.

A general quantum-mechanical formal issue and important practical task of data interpretation is the extraction of the \emph{most complete} quantum-mechanical information from the accessible observables.
Important but very rare case when extraction of the \emph{complete} quantum-mechanical information from data is possible is elastic scattering: from angular distributions one can, in principle, extract set of phase shifts which contains all possible information about this process. For other classes of experimetal data extraction of complete quantum-mechanical information from observables suffers from different types of continuous and discrete ambiguities. For certain classes of reactions the \emph{most complete} quantum-mechanical information which can be extracted is contained in the density matrix.
Because of internal symmetries the density matrix could provide very compact form of data representation depending just on very few parameters. In the case of the pole approximation considered for the $^1$H($^{6}$Li,$^{6}$Be)$n$ reaction there are just four parameters for specific kinematical point: the $0^+/2^+$ ratio, the $0^+/2^+$ relative phase, and two parameters describing $2^+$ state alignment. The density matrix approximation may be questioned for such complicated process as charge-exchange reaction.
Despite this issue we demonstrate in this work principal ability to describe very complex and detailed multi-dimensional correlation patterns by applying this compact formalism.

In the mentioned recent three-body correlation studies, the detailed correlation data allowed to resolve the following intriguing questions.

\noindent (i) It was possible to check consistency of the long-range aspect of the three-body problem in continuum \cite{Grigorenko:2009,Brown:2014}.

\noindent (ii) We were able to figure out fine details of the decay dynamics for democratic decays by examples of $^{6}$Be and $^{16}$Ne ground and first excited states \cite{Grigorenko:2009c,Brown:2014,Brown:2015a}.

\noindent (iii) Possibility to uncover weakly populated states due to interference with ``background'' states was demonstrated in \cite{Golovkov:2004a,Golovkov:2005}.

In this work we add one more point to this list of scientific tasks which can be resolved by correlation studies.
The basic point here is that correlation data are \emph{very detailed} to make possible investigation in different regions of the kinematical space. Such a detailed information is not easily accessible in exotic dripline systems where secondary beams typically have low or modest intensities. Our work provide additional motivation for this type of reaction studies.


\section{Conclusions}


The correlation data for three-body $\alpha$+$p$+$p$ decay of the $^{6}$Be continuum with overlapping states populated in the $^1$H($^{6}$Li,$^{6}$Be)$n$ charge-exchange reaction were analyzed. The energy region \Et$<3$\,MeV, where low-lying $0^+$ and $2^+$ states are populated, has been considered. Experimental data of high statistics ($\sim 5 \times 10^6$ reconstructed events) allowed us to investigate correlations with reasonable resolution both in the $^{6}$Be excitation energy and in the reaction center-of-mass angle. Data analysis was carried out by using the comparison of experimental data with MC simulations taking into account the population of $0^+$ and $2^+$ states in the $^{6}$Be continuum and neglecting the population of $J^-$ continuum. Our treatment showed that internal structure of three-body system with broad overlapping states may be revealed in correlations. While internal correlations are weakly sensitive to the investigated parameters (interference between the $0^+$ and $2^+$ states and alignment of $2^+$ state), we observed strong sensitivity to those parameters in external correlations.

The principal opportunity to extract the density-matrix parameters, characterizing the reaction mechanism of population of the $^{6}$Be states, was demonstrated. The suggested method of analysis allows for identification of such fine effects like the ratio of the populated states, interference between them and alignment of the states with $J$$>$1/2 for other nuclei, and it may be regarded as a general tool for similar tasks.

Nice examples of the employment of the three-body correlations for spin-parity identification are high-statistics experimental file $^{5}$H \cite{Golovkov:2004a,Golovkov:2005}, low-statistics set $^{10}$He \cite{Sidorchuk:2012} and high-precision treatment of the three-body Coulomb continuum effects in $^{16}$Ne \cite{Brown:2014}.
The results obtained in this work provide exemplary demonstration how the high-statistics few-body correlation data can be used for determination of the fine effects of the reaction mechanism. This work underline the importance of the high-statistics studies of the few-body correlations as important point of experimental agenda of RIB facilities.

%
\section{Acknowledgements}
%

This  work was partly supported by Russian Science Foundation (grant No.\ 17-12-01367) and MEYS Projects (Czech Republic) LTT17003 and LM2015049.


\bibliography{all.bib}


\end{document}